\PassOptionsToPackage{dvipsnames}{xcolor}
\documentclass[oneside,11pt,letterpaper,final,leqno,onefignum,onetabnum]{elsarticle}

\journal{Journal of Computational Physics}

\usepackage[american]{babel}
\usepackage[utf8]{inputenc}
\usepackage[T1]{fontenc}

\usepackage{booktabs}
\usepackage{mathtools}
\usepackage{subfig}
\usepackage{color}
\usepackage{comment}
\usepackage{optidef}

\usepackage{amsmath}
\usepackage{amssymb,amsfonts}
\usepackage{amsthm}
\newtheorem{proposition}{Proposition}

\usepackage{graphicx}

\newcommand{\C}{\mathbb{C}}
\newcommand{\e}{\varepsilon}

\newcommand{\oer}{\overline{\varepsilon}_{r}\,}
\newcommand{\omur}{\overline{\mu}_{r}\,}
\newcommand{\er}{\varepsilon_{r}\,}
\newcommand{\eri}{\varepsilon_{r}^{-1}\,}

\newcommand{\te}{\varepsilon\,}
\newcommand{\mur}{\mu_{r}\,}
\newcommand{\muri}{\mu_{r}^{-1}\,}

\renewcommand{\vec}[1]{\boldsymbol{#1}}

\newcommand{\p}{\partial}
\newcommand{\pn}{\partial_\nu}
\newcommand{\pt}{\partial_\tau}

\newcommand{\ssr}{\sigma^\Sigma_r}

\newcommand{\vE}{\vec E}
\newcommand{\vH}{\vec H}

\newcommand{\vn}{\vec \nu}
\newcommand{\vt}{\vec \tau}

\newcommand{\vx}{\vec x}

\newcommand{\RN}[1]{%
  \textup{\uppercase\expandafter{\romannumeral#1}}%
}

\newcommand{\dx}{\,{\mathrm d}x}
\newcommand{\dox}{\,{\mathrm d}o_x}

\newcommand{\js}[1]{\left[#1\right]_{\Sigma}}

\begin{document}

\begin{frontmatter}

  \title{%
    Nonlinear eigenvalue problems for coupled
    Helmholtz equations
    modeling gradient-index graphene waveguides
  }

  \author[minneapolis,muri,nsf1]{Jung Heon Song}
  \ead{songx762@umn.edu}

  \author[collegestation,muri,nsf2]{Matthias Maier\corref{corresponding}}
  \cortext[corresponding]{Corresponding author}
  \ead{maier@math.tamu.edu}

  \author[minneapolis,muri,nsf1]{Mitchell Luskin}
  \ead{luskin@umn.edu}

  \address[minneapolis]{
      School of Mathematics, University of Minnesota, 206 Church Street SE,
    Minneapolis, MN 55455, USA.}
  \address[collegestation]{
      Department of Mathematics, Texas A\&M University, 3368 TAMU,
    College Station, TX 77843, USA.}

  \fntext[muri]{%
    The authors were supported in part by ARO MURI Award W911NF-14-1-0247.}
  \fntext[nsf1]{The first and third author's research was supported in part
    by NSF Awards DMS-1819220 and DMS-1906129.}
  \fntext[nsf2]{This author’s research was supported in part by NSF Award
    DMS-1912847.}

  \begin{abstract}
    We discuss a quartic eigenvalue problem arising in the context of an
    optical waveguiding problem involving atomically thick \emph{2D}
    materials. The waveguide configuration we consider consists of a
    gradient-index (spatially dependent) dielectric equipped with conducting
    interior interfaces. This leads to a quartic eigenvalue problem with
    mixed transverse electric and transverse magnetic modes, and strongly
    coupled electric and magnetic fields. We derive a weak formulation of
    the quartic eigenvalue problem and introduce a numerical solver based
    on a \emph{quadratification approach} in which the quartic eigenvalue
    problem is transformed to a \emph{spectrally equivalent}
    companion problem. We verify our numerical framework against
    analytical solutions for prototypical geometries. As a practical
    example, we demonstrate how an improved quality factor (defined by the
    ratio of the real and the imaginary part of the computed eigenvalues)
    can be obtained for a family of gradient-index host materials with
    internal conducting interfaces. We outline how this result lays the
    groundwork for solving related shape optimization problems.
  \end{abstract}

\end{frontmatter}


\section{Introduction}

Surface plasmon polaritons (SPPs) are charge density waves that are coupled
to electromagnetic (EM) waves at the interface between a metal and a
dielectric substrate. Exhibiting strong confinement and relatively low
propagation losses, they are thought to be a novel way to confine and
control light on the subwavelength scale in the field of nanophotonic
technology. Such SPPs can be excited in graphene, a two-dimensional carbon
allotrope with a single atom layer that is arranged in a honeycomb lattice
structure~\cite{geim04}. It is characterized by strong confinement, low
losses, and extreme tunability~\cite{geim04,bludov13}; In the infrared
regime, the electric surface conductivity of such a 2D material is
characterized by being complex-valued with a dominant positive imaginary
part. This allows for the propagation of SPPs. Plasmons on graphene offer
not only a lower ohmic loss than conventional plasmonic materials, but also
a strong subwavelength confinement of the EM field~\cite{liu16,oulton09}.
The tunability of graphene by electrical gating or chemical doping, makes
graphene a promising candidate for future compact plasmon
devices~\cite{vakil11}.

A conventional approach of analyzing a waveguide problem is to first reduce
Maxwell's equations to a Helmholtz eigenvalue problem. For a homogeneously
filled waveguide, the EM fields decouple from one another, making the
numerical simulation straightforward. However, if spatially dependent
material parameters (gradient-index materials) are introduced, the field
components are no longer independent from each other, and we are left with
a coupled \emph{nonlinear eigenvalue problem}.

Computational approaches for solving nonlinear eigenvalue problems have
been studied in the literature \cite{gavin18,jay19,jay20,bai18}. They often
require specialized solvers not readily available in current finite element
toolkits~\cite{dealII91}. In this paper we pursue a different approach that
allows to use well established, existing linear algebra techniques for
solving linear eigenvalue problems. To that end, we investigate a general
class of waveguide configurations that consist of spatially dependent
material parameters and contain (arbitrarily shaped) interior conducting 2D
material interfaces. In detail, our contributions are as follows:

\begin{itemize}
  \setlength\itemsep{0.25em}
  \item[--]
    We derive a variational, nonlinear quartic eigenvalue problem for a
    waveguiding problem incorporating spatially dependent material
    parameters and interior conducting interfaces (see Section
    \ref{subsec:variational}). The nonlinear quartic character of the
    eigenvalue problem stems from the fact that the spatially dependent
    material parameters cause a strong coupling between the otherwise
    decoupled transverse magnetic (TM) and transverse electric (TE) modes
    (as would normally be the case for the Helmholtz equation).
  \item[--]
    We solve the quartic eigenvalue problem numerically by transforming it
    to a spectrally equivalent companion problem using a
    quadratification~\cite{teran14} approach. Additional numerical tools,
    such as the M{\"o}bius transform and a perfectly matched layer (PML)
    are employed to assist with solving the eigenvalue problem. We verify
    our numerical method against analytical solutions for prototypical
    geometries with internal conducting interfaces.
  \item[--]
    As a practical example, we demonstrate how an improved quality factor
    (defined by the ratio of the real and the imaginary part of the
    computed eigenvalues) can be obtained (a) for a family of
    gradient-index host materials, and (b) by deformation of the geometry
    (see Section \ref{subsec:results}). Finally, we outline how our
    framework lays the groundwork for solving related shape optimization
    problems.
\end{itemize}
%


\subsection{Related works}

Optical properties of cylindrical waveguides with graphene interfaces have
been extensively studied in the engineering
community~\cite{liu16,xu18,gao14a,gao14b}. Recently, focus has also shifted
to gradient-index structures that couple with
graphene~\cite{xu16,moharrami20}. These structures are based on planar and
cylindrical graphene-dielectric multilayer metamaterials, and have shown
potential applications in terahertz imaging, sensing, detecting, and
communication areas~\cite{xu16,moharrami20}.
In addition, the optimal design of \emph{graded-refractive index
antireflection coatings} has been investigated~\cite{zhang12}. The
motivation behind our work is to formulate a numerical framework that is
specifically designed for solving optical waveguiding problems with
spatially dependent material parameters.

Numerical methods that compute eigenvalues of inhomogeneously loaded
domains have been described before~\cite{dai13a,dai13b,chew99,white02}. For
example, a finite difference frequency-domain method is used to analyze
eigenmodes of inhomogeneously loaded rectangular waveguides
in~\cite{dai13a}. Another study~\cite{white02} presents a method for
computing solenoidal eigenmodes and the corresponding eigenvalues of the
vector Helmholtz equation. We point out that a structurally similar
nonlinear eigenvalue problem also arises in the context of quantum
transmission problems described by the Schr\"odinger
equation~\cite{shao95}: There, a fourth-order eigenvalue problem
\cite[Eq.~(37)]{shao95} emerges for a wave number that is then solved
numerically by a linearized companion problem \cite[Eq.~(39)]{shao95}.
While similar in the resulting linear algebra structure to our waveguiding
problem, neither of the above references directly address the question of
eigenvalue problems with lower-dimensional conducting interfaces.

There exist a number of numerical methods for directly computing
approximations of nonlinear eigenvalue problems. For example, the FEAST
algorithm~\cite{polizzi09,gavin18} uses complex contour integration to
compute a cluster of eigenvalues within some user-defined region in the
complex plane. As such it is also well suited to compute solutions of
quartic eigenvalue problems. It has been successfully used for simulating
the propagation of light through optical fibers~\cite{jay19,jay20}. Another
numerical computing technique is based on the equivalent Rayleigh quotient
optimization problem~\cite{bai18}. Here, a nonlinear eigenvalue problem is
solved using a spectral transformation based on nonlinear shifting and a
reformulation using second-order derivatives. In addition, an increasing
number of mathematical software packages, such as for example SLEPc and the
Julia programming language, provide black box solvers for polynomial
eigenvalue problems~\cite{slepc05,bezanson17}.


\subsection{Paper organization}

The remainder of the paper is organized as follows. In Section
\ref{sec:variational}, we derive a variational quartic eigenvalue problem
for the waveguiding problem based on time-harmonic Maxwell's equations.
In Section~\ref{sec:numerics}, we describe our numerical approach for
solving the quartic eigenvalue problem, including a linearization based on
quadratification, the use of a M{\"o}bius transform to shift the spectrum,
and a PML.
Section~\ref{sec:validation} discusses and derives analytical solutions for
prototypical configurations, which will be used to verify our numerical
method in the subsequent section.
Section~\ref{sec:results} presents numerical results in domains
with and without azimuthal symmetry. We demonstrate how our numerical
method can be extended to arbitrary computational domains.
Section~\ref{sec:conclusion} concludes the paper with a summary of our
results and an outlook.


\section{Variational formulation}
\label{sec:variational}
We introduce a variational formulation for a relevant eigenvalue
problem prescribed with a gradient-index host material with (arbitrarily shaped)
conducting interfaces in the context of a waveguide configuration.
A convenient rescaling of the equations to dimensionless forms is
used~\cite{maier17a}.


\subsection{Preliminaries}

\begin{figure}[t]
  \centering
    \vspace{-2em} 
    \subfloat[]{
      \resizebox{5.5cm}{!}{
        \includegraphics{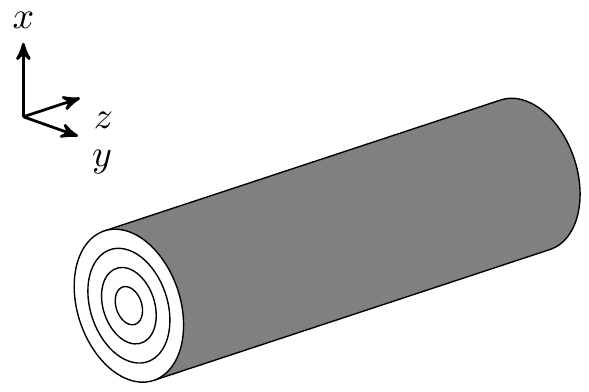}}}
    \hspace{2em}
    \subfloat[]{
    \resizebox{!}{4.2cm}{
    \includegraphics{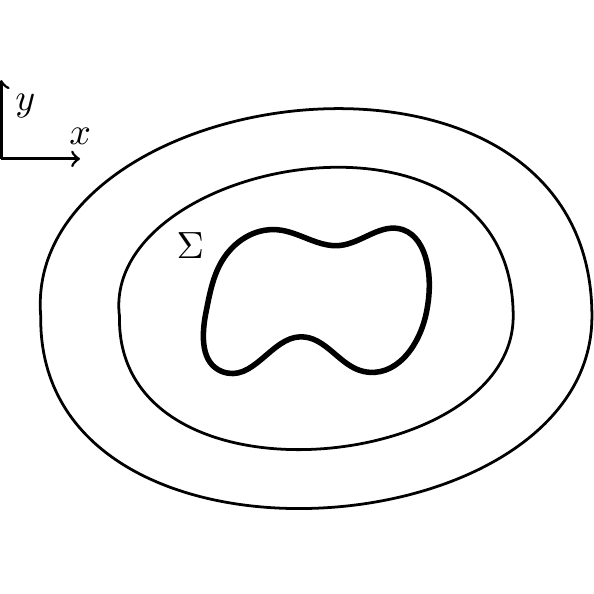}}}

  \caption{
    (a) Schematic of a prototypical multilayer waveguide. (b)
    Cross-sectional schematic of the computational domain.
    The closed curve, $\Sigma$, is prescribed with a non-vanishing conductivity.
    The waveguide is characterized by material parameters $\te(x,y)$ and $\mu(x,y)$.
  }
  \label{fig:schematicdomain}
\end{figure}
The source-free time-harmonic Maxwell's equations are given by
\begin{align}
  \begin{cases}
    \begin{aligned}
      \nabla\times\vE \;&\;= i\omega\mu \vH, \\
      \nabla\times\vH \;&\;= -i\omega\te\vE,
    \end{aligned}
  \end{cases}
  \label{eq:unscaled_maxwell}
\end{align}
where $\vE(\vx)$ and $\vH(\vx)$ denote the electric and magnetic field,
respectively, and $\omega$ is the temporal frequency. $\mu(x,y)$ and
$\te(x,y)$ are complex-valued functions of the transverse coordinates,
where $\mu$ denotes the magnetic permeability and $\te$ denotes the
electric permittivity (see Figure~\ref{fig:schematicdomain}). In order to
study \emph{guided modes} we make the additional ansatz
\begin{align*}
  \vec{\mathcal{F}} \;\sim\; e^{i\,k_z\,z},
\end{align*}
and decompose the fields $\vec{\mathcal{F}}=\{\vE,\vH\}$ and the gradient
operator, $\nabla$, into their longitudinal and transverse parts, whence we
obtain
\begin{align*}
  \vec{\mathcal{F}}=\vec{\mathcal{F}}_s+\hat{z}\mathcal{F}_z,
  \quad \nabla=\nabla_s + ik_z\hat{z},
\end{align*}
where the subscript $\emph{s}$ denotes the transverse direction and $\hat
z$ denotes the unit vector in $z$-direction. In the strong sense,
(\ref{eq:unscaled_maxwell}) holds true everywhere except on the points
comprising the conducting interface, $\Sigma$. The surface conductivity
$\sigma^\Sigma(x, y)$ on the conducting interface $\Sigma$ gives rise to a jump
condition on the tangential part of the magnetic field \cite{maier17a}. In
summary, we obtain
\begin{align}
  \begin{cases}
    \begin{aligned}
      \left[\vn\times(\vH_s + \hat{z}H_z)\right]_\Sigma & = \sigma^\Sigma
      \big(\vn\times(\vE_s + \hat{z}E_z)\big)\times\vn,
      \\
      \left[\vn\cdot\mu(\vH_s + \hat{z}H_z)\right]_\Sigma & = 0,
      \\
      [\vn\times(\vE_s + \hat{z}E_z)]_\Sigma  &= 0,
      \\
      [\vn\cdot\te(\vE_s + \hat{z}E_z)]_\Sigma &= \frac{1}{i\omega}\nabla\cdot(\sigma^\Sigma\vE),
    \end{aligned}
  \end{cases}
  \label{eq:unscaled_jump}
\end{align}
where $\vn$ is a chosen unit normal vector field on $\Sigma$, and
$\js{\,.\,}$ denotes the jump over $\Sigma$ with respect to $\vn$, viz.,
\begin{align*}
  \js{\vec F}(\vx)
  \,:=\,
  \lim_{\alpha\searrow0}
  \Big(\vec F(\vx+\alpha\vn) - \vec F(\vx-\alpha\vn)\Big)\qquad
  \vx\in\Sigma.
\end{align*}
We also fix the notation
\begin{align*}
  \vec F^+ \,:=\, \lim_{\alpha\searrow0} \vec F(\vx+\alpha\vn),
  \quad
  \vec F^- \,:=\, \lim_{\alpha\searrow0} \vec F(\vx-\alpha\vn),
  \quad \text{for}\;\vx\in\Sigma.
\end{align*}
Next we introduce a convenient rescaling of the system to dimensionless
form by setting the characteristic wavenumber of the ambient space to
1~\cite{maier17a}:
\begin{gather*}
  \vx \;\rightarrow\; \breve\vx = k_0\vx,
  \quad
  \nabla_s \;\rightarrow\; \breve\nabla_s = \frac1{k_0}\,\nabla_s,
  \\
  \mu \;\rightarrow\; \mur =\frac1{\mu_0}\mu,
  \quad
  \e \;\rightarrow\; \er = \frac1{\e_0}\te,
  \quad
  \sigma^\Sigma \;\rightarrow\; \ssr =
  \sqrt{\frac{\mu_0}{\e_0}}\,\sigma^\Sigma
  \\
  \vE \;\rightarrow\; \breve{\vE} = \frac{k_0^2}{\omega\mu_0}\vE,
  \quad
  \vH \;\rightarrow\; \breve{\vH} = k_0\vH,
  \quad
  k_z \;\rightarrow\; \breve{k}_z = \frac{k_z}{k_0}.
\end{gather*}
To lighten the notation, we omit the breve sign in the remainder of
this paper. Applying the rescaling to \eqref{eq:unscaled_jump} and
rewriting into tangential and normal parts leads to the following interface
conditions:
\begin{align}
  \begin{cases}
    \begin{aligned}
      [\vH_s]_\Sigma\cdot\vt
      \;&=&\;\left[\frac{i}{k_s^2} \left(k_z\pt H_z + \er\pn E_z\right)\right]_\Sigma
      \;&=\;\ssr E_z,
      \\[0.2em]
      & & [H_z]_\Sigma \;&=\; -\ssr\vE_s\cdot\vt,
      \\[0.2em]
      [\mur\vH_s]_\Sigma\cdot\vn
      \;&=&\;\left[\frac{i\mur}{k_s^2} \left(k_z\pn H_z - \er\pt E_z\right)\right]_\Sigma
      \;&=\;0,
      \\[0.2em]
      [\vE_s]_\Sigma\cdot\vt
      \;&=&\;\left[\frac{i}{k_s^2} \left(k_z\pt E_z - \mur\;\pn H_z\right)\right]_\Sigma
      \;&=\;0,
      \\[0.2em]
      & & [E_z]_\Sigma \;&=\; 0,
      \\[0.2em]
      [\er\vE_s]_\Sigma\cdot\vn
      \;&=&\;\left[\frac{i\er}{k_s^2} \left(k_z\pn E_z + \mur\,\pt H_z\right)\right]_\Sigma
      \;&=\;\frac1i\;\nabla_s\cdot(\ssr\vE_s),
    \end{aligned}
  \end{cases}
  \label{eq:jump}
\end{align}
where $\pt$ and $\pn$ denote the derivative in the tangential and the
normal direction, respectively; $k_s(x,y)^2 = \mur(x,y)\;\er(x,y) - k_z^2$
is a function in the transverse direction; and where we have used the
identities ({see Appendix~\ref{app:derivation}})
\begin{align}
  \begin{cases}
    \begin{aligned}
      k_s^2\vE_s & = i\left(k_z\nabla_s E_z+\mur\nabla_s
      \times H_z\right), \\[0.2em]
      k_s^2\vH_s & = i\left(k_z\nabla_s H_z-\er\nabla_s
      \times E_z\right).
    \end{aligned}
  \end{cases}
  \label{eq:rescaled_transverse}
\end{align}
The first-order system (\ref{eq:unscaled_maxwell}) can be manipulated in a
similar fashion ({see \ref{app:derivation}}) to obtain
\begin{align}
  \begin{cases}
  \begin{aligned}
    & \nabla_s\times\left(\muri\nabla_s\times\hat{z}E_z\right) +
      ik_z\nabla_s\times\left(\muri\hat{z}\times\vE_s\right)-\er E_z\;=\;0, \\[0.2em]
    & \nabla_s\times\left(\eri\nabla_s\times\hat{z}H_z\right) +
      ik_z\nabla_s\times\left(\eri\hat{z}\times\vH_s\right)+\mur H_z\;=\;0.
  \end{aligned}
  \end{cases}
  \label{eq:secondorder_rescaled}
\end{align}
%


\subsection{Variational Statement}
\label{subsec:variational}

Let $\Omega\subset\mathbb{R}^n$, where $n = 2,3$, be a simply connected and
bounded domain with Lipschitz-continous piecewise smooth boundary,
$\p\Omega$. Assume, in addition, that $\Sigma$ is a Lipschitz-continuous,
piecewise smooth hypersurface. Let $\vn$ and $\vt$ denote the outer normal
and the tangential vector on $\Sigma$ (see
Figure~\ref{fig:template_domain}). Some algebraic manipulation shows that
$\nabla_s\times\left(\muri\nabla_s\times\hat{z}E_z\right)\;=\;
      -\nabla_s\cdot\left(\muri\nabla_s\hat{z}E_z\right)$ and
$\nabla_s\times\left(\muri\hat{z}\times\vE_s\right)\;=\;
      \hat{z}\nabla_s\cdot\left(\muri\vE_s\right)$,
which can be used in conjunction with (\ref{eq:rescaled_transverse}) and
(\ref{eq:secondorder_rescaled}) to obtain:
\begin{align}
  \begin{cases}
  \begin{aligned}
    -\nabla_s\cdot\left(\frac{\er}{k_s^2}\nabla_sE_z\right) -
      k_z\nabla_s\cdot\left(\frac1{k_s^2}\nabla_s\times\hat{z}H_z\right) -
      \er E_z\;&=\;0, \\
    -\nabla_s\cdot\left(\frac{\mur}{k_s^2}\nabla_sH_z\right) +
      k_z\nabla_s\cdot\left(\frac1{k_s^2}\nabla_s\times\hat{z}E_z\right) -
      \mur H_z\;&=\;0.
  \end{aligned}
  \end{cases}
  \label{eq:rescaled_z}
\end{align}
We observe that if the domain is homogeneously filled and isotropic, the curl
terms vanish, yielding the familiar decoupled Helmholtz equation for $E_z$ and
$H_z$. Now assume that
\begin{align}
  [\er]_\Sigma \;=\; 0,
  \qquad
  [\mur]_\Sigma \;=\; 0.
  \label{eq:assumptions}
\end{align}
For the sake of brevity, we summarize the derivation here and refer the
reader to \ref{app:derivation} for details. We now want to remove the
$k_s^2$ term in the denominator. Because $k_s(x,y)$ is a spatially
dependent function, the lowest power of $k_s$ that achieves this goal is
$k_s^4$. Thus, by multiplying (\ref{eq:rescaled_z}) with $k_s^4$ and
testing the first equation by $\varphi$ and the second equation by $\psi$,
we obtain
\begin{multline}
  (\mur\varepsilon_r^2\,\nabla_s E_z,\nabla_s\varphi)
    +2(\er\nabla_s E_z,\nabla_s(\oer\omur)\varphi) \\
      +k_z(\mur\er\nabla_s\times\hat{z}H_z,\nabla_s\varphi)
      -k_z^2(\er\nabla_s E_z,\nabla_s\varphi) \\
      +2k_z(\nabla_s\times\hat{z}H_z,\nabla_s(\oer\omur)\varphi)
      -k_z^3(\nabla_s\times\hat{z}H_z,\nabla_s\varphi) - (\er k_s^4 E_z,\varphi) \\
  + (\er\mu_r^2\,\nabla_s H_z,\nabla_s\psi)+2(\mur\nabla_s H_z,\nabla_s(\oer\omur)\psi) \\
      -k_z(\mur\er\nabla_s\times\hat{z}E_z,\nabla_s\psi)
      -k_z^2(\mur\nabla_s H_z,\nabla_s\psi) \\
      -2k_z(\nabla_s\times\hat{z}E_z,\nabla_s(\oer\omur)\psi)
      +k_z^3(\nabla_s\times\hat{z}E_z,\nabla_s\psi) -(\mur k_s^4 H_z,\psi) \\
  -i\langle \ssr k_s^4 E_z,\varphi\rangle_\Sigma \;=\; 0.
  \label{eq:weakform_ugly}
\end{multline}
This shows the following statement.
\begin{proposition}
  \label{prop:quartic_problem}
  Provided that $k_s\neq0$ and $[\er]_\Sigma = [\mur]_\Sigma=0$,
  the nonlinear eigenvalue problem
  \begin{align}
    \tag{N}
    \text{Find $k_z\in\C\backslash\{0\}$ and $E_z$, $H_z$ s.\,t.
    \eqref{eq:rescaled_z} and \eqref{eq:jump} are satisfied}
    \label{eq:nonlinear}
  \end{align}
  can be restated as a quartic eigenvalue problem
  \begin{multline}
    \label{eq:quartic} \tag{Q}
    \text{Find $k_z\in\C\backslash\{0\}$ and $E_z$, $H_z\in \vec X(\Omega)
      = \{(u,v): u,v\in H^1(\Omega,\mathbb{C})\}$ s.\,t.}
    \\[0.25em]
    \mathcal{Q}\big(k_z, (E_z,H_z)\big)\big[(\varphi,\psi)\big] \;=\; 0
    \qquad\text{for all $\varphi$, $\psi\in H^1(\Omega,\mathbb{C})$,}
    \notag
  \end{multline}
  where
  \begin{align*}
    \mathcal{Q}\big(k_z,(E_z,H_z)\big)\big[(\varphi,\psi)\big] \;=\;
    \sum_{l=0}^4 (k_z)^l\,a_l\big((E_z,H_z)\big)\big[(\varphi,\psi)\big],
  \end{align*}
  and
  \begin{align}
    \begin{aligned}
      \begin{cases}
        a_0((E_z,H_z),(\varphi,\psi)) =
        (\mur\varepsilon_r^2\nabla_s E_z,\nabla_s\varphi)
          +2(\er\nabla_s E_z,(\nabla_s\omur\oer)\varphi) \\[0.2em]
          \qquad\qquad\qquad\qquad\qquad\quad
          -(\mu_r^2\,\varepsilon_r^3\,E_z,\varphi)
          + (\er\mu_r^2\,\nabla_s H_z,\nabla_s\psi) \\[0.2em]
          \qquad\qquad\qquad\qquad\qquad\qquad
          +2(\mur\nabla_s H_z,(\nabla_s\omur\oer)\psi) \\[0.2em]
          \qquad\qquad\qquad\qquad\qquad\qquad\quad
          -(\mu_r^3\,\varepsilon_r^2\,H_z,\psi)
          -i\langle\ssr\mu_r^2\,\varepsilon_r^2\,E_z,\varphi\rangle_\Sigma,
        \\[0.5em]
        a_1((E_z,H_z),(\varphi,\psi)) =
          (\mur\er\nabla_s\times\hat{z}H_z,\nabla_s\varphi) \\[0.2em]
          \qquad\qquad\qquad\qquad\qquad\quad
            + 2(\nabla_s\times\hat{z}H_z,(\nabla_s\omur\oer)\varphi) \\[0.2em]
          \qquad\qquad\qquad\qquad\qquad\qquad\quad
          - (\mur\er\nabla_s\times\hat{z}E_z,\nabla_s\psi) \\[0.2em]
          \qquad\qquad\qquad\qquad\qquad\qquad\qquad\quad
            - 2(\nabla_s\times\hat{z}E_z,(\nabla_s\omur\oer)\psi)
        \\[0.5em]
        a_2((E_z,H_z),(\varphi,\psi)) =
          -(\er\nabla_s E_z,\nabla_s\varphi) +
          2(\mur\varepsilon_r^2\,E_z,\varphi)
          \\[0.2em]
          \qquad\qquad\qquad\qquad\qquad\quad
          - (\mur\nabla_s H_z,\nabla_s\psi) + 2(\mu_r^2\,\er H_z,\psi) \\[0.2em]
          \qquad\qquad\qquad\qquad\qquad\qquad\quad
          +2i\langle\ssr\mur\er E_z,\varphi\rangle_\Sigma,
        \\[0.5em]
        a_3((E_z,H_z),(\varphi,\psi)) =
          - (\nabla_s\times\hat{z}H_z,\nabla_s\varphi)
          + (\nabla_s\times\hat{z}E_z,\nabla_s\psi),
        \\[0.5em]
        a_4((E_z,H_z),(\varphi,\psi))
          = -(\er E_z,\varphi) - (H_z,\psi)- i\langle\ssr
          E_z,\varphi\rangle_\Sigma.
      \end{cases}
    \end{aligned}
    \label{eq:bilinear_forms}
  \end{align}
\end{proposition}


\section{Numerical approach}
\label{sec:numerics}

In this section, we outline our numerical approach for solving the quartic
eigenvalue problem~\eqref{eq:quartic}. In particular we discuss a
\emph{quadratification} approach transforming the quartic eigenvalue
problem into a companion problem with equivalent spectrum. A perfectly
matched layer (PML), an artificial sponge layer placed near the boundary
such that all outgoing waves decay exponentially, is introduced. The
variational formulation~\eqref{eq:quartic} is discretized on a non-uniform
quadrilateral mesh.
\begin{proposition}
  Let $\vec X_h(\Omega) \subset \vec X(\Omega)$ be a finite element
  subspace spanned by Lagrange finite elements $Q_p$.
\end{proposition}
\begin{align}
  \tag{Q$_h$}
  \sum_{l=0}^4 k_z^l a_l((E_z,H_z),(\varphi,\psi)) = 0,
  \label{eq:quartic_discretized}
\end{align}
Our goal is to translate~\eqref{eq:quartic_discretized} into a finite
dimensional linear problem, which then allows the use of a standard linear
algebra solver.


\subsection{Construction of a companion problem to the quartic eigenvalue problem}

We build upon the algebraic tool of \emph{quadratification} introduced
in~\cite{teran14}, which allows us to reduce any even power matrix
polynomial eigenvalue problem to a spectrally equivalent linear eigenvalue
problem. Prop.~\ref{prop:quadratification} summarizes the main result (for
a more general discussion of the ideas behind this reduction procedure,
we refer the reader to~\cite{teran14, drmac19}).
\begin{proposition}[Theorem 5.3 and 5.4 of \cite{teran14,drmac19}]
  \label{prop:quadratification}
  Consider a quartic eigenvalue problem, to find $\lambda\in\C$, and
  $x\in\C^n$, s.\,t.
  \begin{align*}
    \sum_{k=0}^4 \lambda^k A_k x = 0,
  \end{align*}
  where $A_k\in\C^{n\times n}$ are given matrices. Then, the
  \emph{linearization} stated below is spectrally equivalent to the
  original problem (c.f.~\cite{teran14} Theorem 5.3 and 5.4): Find
  $\lambda\in\C$, and $z\in\C^{4n}$ s.\,t.
  \begin{align}
    \begin{pmatrix}
      A_3 &    0 & -I_n &    0 \\
      A_1 &    0 &    0 & -I_n \\
      0   & -I_n &    0 &    0 \\
      A_0 &    0 &    0 &    0
    \end{pmatrix}z
    \;+\;
    \lambda
    \begin{pmatrix}
      A_4 & 0   & 0 & 0 \\
      A_2 & I_n & 0 & 0 \\
        0 & 0 & I_n & 0 \\
        0 & 0 & 0 & I_n
    \end{pmatrix}z = 0.
  \end{align}
  Here, $I_n$ denotes the $n\times n$ identity matrix.
\end{proposition}

With this result at hand, we rewrite \eqref{eq:quartic_discretized} as a
linear eigenvalue problem:
\begin{align}
  \begin{aligned}
    S\,z + \lambda M z \;=\; 0,
  \end{aligned}
  \tag{LQ$_h$}
  \label{eq:quadratification}
\end{align}
where
\begin{align*}
  {S} = \begin{pmatrix}
      a_3 & 0 & -I_n & 0 \\
      a_1 & 0 & 0 & -I_n \\
      0 & -I_n & 0 & 0   \\
      a_0 & 0 & 0 & 0
    \end{pmatrix},
  \quad
  {M} = \begin{pmatrix}
      a_4 & 0 & 0 & 0 \\
      a_2 & I & 0 & 0 \\
      0 & 0 & I & 0   \\
      0 & 0 & 0 & I
  \end{pmatrix},
  \quad
  {z} = \begin{pmatrix}z_1 \\ z_2 \\ z_3 \\ z_4 \end{pmatrix}.
\end{align*}
Here, by some abuse of notation $a_i$ denotes the corresponding matrix
formed by the bilinear form $a_i(.\,.)$ given in \eqref{eq:bilinear_forms}
and by fixing a basis of $\vec X_h(\Omega)$. A quick computation shows that
the eigenvectors of the original problem (\ref{eq:quartic_discretized}) and
of the final linearized problem (\ref{eq:quadratification}) are related as
follows.

\begin{proposition}
  Let $\lambda\in\C$ and $x\in\C^n$ be an eigenvalue with corresponding
  eigenvector of \eqref{eq:quartic_discretized}. Then, $\lambda$ and
  $z\in\C^{4n}$ given by
  \begin{align}
    \begin{cases}
      z_1 \;&=\; \lambda x,
      \\
      z_2 \;&=\; \lambda^2 (a_3+\lambda a_4)x,
      \\
      z_3 \;&=\; \lambda   (a_3+\lambda a_4)x,
      \\
      z_4 \;&=\; -a_0 x,
    \end{cases}
    \label{eq:characterization_z}
  \end{align}
  is an eigenvalue with corresponding eigenvector of
  \eqref{eq:quadratification}. Conversely, if $\lambda\in\C$ and
  $z\in\C^{4n}$ is an eigenvalue and eigenvector pair of
  \eqref{eq:quadratification}, then provided that $\det(a_0)\not=0$ and
  $\det(\lambda a_4+a_3)\not=0$, the vector $z$ is characterized by
  \eqref{eq:characterization_z} and $\lambda$ and $x$ are an eigenvalue and
  eigenvector pair of \eqref{eq:quartic_discretized}.
\end{proposition}


\subsection{Perfectly Matched Layer}
\label{subsec:pml}

\begin{figure}[t]
  \centering
  \includegraphics{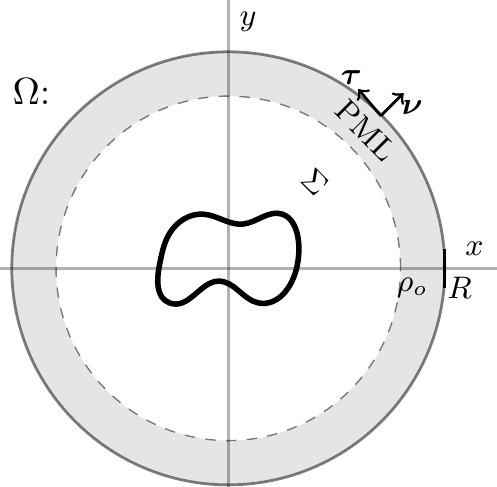}
  \caption{
    Schematic of the computational domain of a circular waveguide
    $\Omega$, with boundary $\p\Omega$, outer normal vector $\vn$, and
    tangential vector $\vt$. A conducting ciruclar interface is labeled
    $\Sigma$. A
    perfectly matched layer (PML) is enforced in the shaded region.}
  \label{fig:template_domain}
\end{figure}
A perfectly matched layer (PML) is a truncation procedure motivated from
electromagnetic scattering problems in the time domain. The idea is to
surround the computational domain with a so-called sponge layer, an
artificial boundary wherein all outgoing electromagnetic waves decay
exponentially with minimal artificial reflection (see
Figure~\ref{fig:template_domain}). As outlined in
\cite{collino98,monk03,maier17a}, we carry out a change of coordinates from
the computational domain with real-valued coordinates to a domain with
complex-valued coordinates. We refer the reader to \cite{collino98} for
details. For a spherical absorption layer, we define the transformation
$\tilde\rho = \rho\overline{d}$, where
\begin{align*}
  & d = 1 + is(r), \quad \overline{d} = 1+i/r\int_\rho^r s(\tau)\,d\tau,
\end{align*}
and $s(\tau)$ is an appropriately chosen, nonnegative scaling function.
Applying the above transformation, the quartic eigenvalue problem takes the
following rescaled form within the PML:
\begin{multline*}
  \hat\nabla_s\cdot(k_s^2(\er\hat\nabla_s\hat E_z))
  - 2k_s^2\hat\nabla_s\cdot(\er\hat\nabla_s\hat E_z)
  + k_z \hat\nabla_s\cdot(k_s^2\hat\nabla_s\times\hat H_z)
  \\
  -2k_zk_s^2\hat\nabla_s\cdot(\hat\nabla_s\times\hat H_z)
  +\hat\nabla_s\cdot(k_s^2(\mur\hat\nabla_s\hat H_z))
  - 2k_s^2\hat\nabla_s\cdot(\mur\hat\nabla_s\hat H_z)
  \\
  + k_z \hat\nabla_s\cdot(k_s^2\hat\nabla_s\times\hat E_z)
  +2k_zk_s^2\hat\nabla_s\cdot(\hat\nabla_s\times\hat E_z) \\
  - \er k_s^4\hat E_z - \mur k_s^4\hat H_z \;=\; 0.
\end{multline*}
This can be rewritten as
\begin{multline*}
  \frac1{d\overline{d}}\nabla_s\cdot(k_s^2\er A\nabla_s E_z)
  - \frac{2k_s^2}{d\overline{d}}\nabla_s\cdot(\er A\nabla_s E_z)
  + \frac{k_z}{d\overline{d}}\nabla_s\cdot(k_s^2\nabla_s\times H_z) \\
  -\frac{2k_zk_s^2}{d\overline{d}}\nabla_s\cdot(\nabla_s\times H_z)
  +\frac1{d\overline{d}}\nabla_s\cdot(k_s^2\er A\nabla_s E_z)
  - \frac{2k_s^2}{d\overline{d}}\nabla_s\cdot(\er A\nabla_s E_z)
  \\
  + \frac{k_z}{d\overline{d}}\nabla_s\cdot(k_s^2\nabla_s\times H_z)
  +\frac{2k_zk_s^2}{d\overline{d}}\nabla_s\cdot(\nabla_s\times E_z) \\
  + \er k_s^4 E_z + \mur k_s^4 H_z = 0,
\end{multline*}
where $A = T_{\vec e_x, \vec e_r}^{-1}\text{diag}(\overline{d}/d,
d/\overline{d})T_{\vec e_x, \vec e_r}$, and $T_{\vec e_x, \vec e_r}$ is the
rotation matrix that rotates $\vec e_r$ onto $\vec e_x$. We enforce the
condition that the material parameters are constant outside the PML, i.e.,
$\er$ and $\mur$ do not undergo a change of coordinate. Additionally,
because the eigenmodes of our interest are confined to the conducting
interface, which is situated inside the PML, no coordinate change is needed
for the jump condition.
The modified bilinear
forms, $\tilde{a}_i$, are
\begin{align}
  \begin{aligned}
    \begin{cases}
      \tilde{a}_0((E_z,H_z),(\varphi,\psi)) =
      (\varepsilon_r^2A\,\nabla_s E_z,\nabla_s\varphi)
        +2(\er A\nabla_s E_z,(\nabla_s\oer)\varphi) \\[0.2em]
        \qquad\qquad\qquad\qquad\qquad\quad
        -d\overline{d}\,(\varepsilon_r^3 E_z,\varphi)
        + (\er A\nabla_s H_z,\nabla_s\psi) \\[0.2em]
        \qquad\qquad\qquad\qquad\qquad\qquad
        +2(A\nabla_s H_z,(\nabla_s\oer)\psi)
        -d\overline{d}\,(\varepsilon_r^2 H_z,\psi) \\[0.2em]
        \qquad\qquad\qquad\qquad\qquad\qquad\quad
        -i\langle\ssr\varepsilon_r^2 E_z,\varphi\rangle_\Sigma,
      \\[0.5em]
      \tilde{a}_1((E_z,H_z),(\varphi,\psi)) =
        (\er\nabla_s\times\hat{z}H_z,\nabla_s\varphi) \\[0.2em]
        \qquad\qquad\qquad\qquad\qquad\quad
          + 2(\nabla_s\times\hat{z}H_z,(\nabla_s\oer)\varphi) \\[0.2em]
        \qquad\qquad\qquad\qquad\qquad\qquad\quad
        - (\er\nabla_s\times\hat{z}E_z,\nabla_s\psi) \\[0.2em]
        \qquad\qquad\qquad\qquad\qquad\qquad\qquad\quad
          - 2(\nabla_s\times\hat{z}E_z,(\nabla_s\oer)\psi)
      \\[0.5em]
      \tilde{a}_2((E_z,H_z),(\varphi,\psi)) =
        -(\er A\nabla_s E_z,\nabla_s\varphi) +
        2d\overline{d}\,(\varepsilon_r^2\,E_z,\varphi)
        \\[0.2em]
        \qquad\qquad\qquad\qquad\qquad\quad
        - (A\nabla_s H_z,\nabla_s\psi) + 2d\overline{d}\,(\er H_z,\psi) \\[0.2em]
        \qquad\qquad\qquad\qquad\qquad\qquad\quad
        +2i\langle\ssr\er E_z,\varphi\rangle_\Sigma,
      \\[0.5em]
      \tilde{a}_3((E_z,H_z),(\varphi,\psi)) =
        - (\nabla_s\times\hat{z}H_z,\nabla_s\varphi)
        + (\nabla_s\times\hat{z}E_z,\nabla_s\psi),
      \\[0.5em]
      \tilde{a}_4((E_z,H_z),(\varphi,\psi))
        = -d\overline{d}\,(\er E_z,\varphi) - d\overline{d}\,(H_z,\psi)- i\langle\ssr
        E_z,\varphi\rangle_\Sigma.
    \end{cases}
  \end{aligned}
  \label{eq:bilinear_forms_pml}
\end{align}
%


\subsection{M{\"o}bius Transform}

Numerical computations of Maxwell eigenvalue problems, in particular with a
perfectly matched layer, often contain a large number of spurious
eigenvalues. Spurious modes are numerical solutions of the vector wave
equation that convey no physical meaning. A number of different approaches
have been proposed to eliminate part or all of the spurious modes, e.g., by
enforcing the solenoidal nature of the flux~\cite{konrad76}, by adding a
penalty factor~\cite{coulomb81, rahman84}, or by solving Maxwell's
equations via the method of constraints~\cite{webb88, konrad85,
kobelansky86}. A readily implementable approach that best suits our
computational setup is through the Mobius transformation, which shifts the
spectrum in such a way that the modes of interest are close to the origin.
They can then be selectively computed with conventional Krylov-space
iteration techniques. The M{\"o}bius transform is a conformal mapping,
defined as follows.
\begin{align*}
  \lambda\mapsto \frac{a\lambda+b}{c\lambda+d},
\end{align*}
where $a,b,c,d\in\mathbb{C}$ are chosen parameters. Over arbitrary fields,
the M{\"o}bius transformation preserves a number of spectral features of
matrix polynomials, such as regularity, rank, minimal indicies, the
location of zero entries, symmetry, and skew-symmetry~\cite{mackey14}. In
particular, every M{\"o}bius transformation preserves the relation of
spectral equivalence~\cite{mackey14}. The computational eigenvalue problem,
after introducing a PML, truncating the domain, and applying a finite
element disretization, can be written as
\begin{align}
  Sz = k_z M z,
\end{align}
for a complex-valued vector $z$ and appropriate complex-valued matrices $S$ and $M$.
The implementation of the PML discussed in \ref{subsec:pml} necessitates
changes to the definition of $a_1, a_2, a_3$, and $a_4$.
The idea is to use the M{\"o}bius transform to map points near the origin
to target values
\begin{align}
  (aS+bM)z = \tilde{k}_z (cS+dM) z,
\end{align}
where $a,b,c,d$ are the M{\"o}bius transform parameters. The original
eigenvalue can be retrieved via the inverse M{\"o}bius transform $k_z =
(-b+d\tilde{k}_z)/(a-c\tilde{k}_z)$.


\section{Validation of weak formulation}
\label{sec:validation}

In this section, the analytical solution for constant material
parameters is derived and discussed. We use the analytic result to validate
our numerical approach.

\begin{figure}[h]
  \centering
  \subfloat[$m = 1$]{
    \includegraphics[width=4.2cm,trim=9.9cm 2cm 7.0cm 4cm,clip]{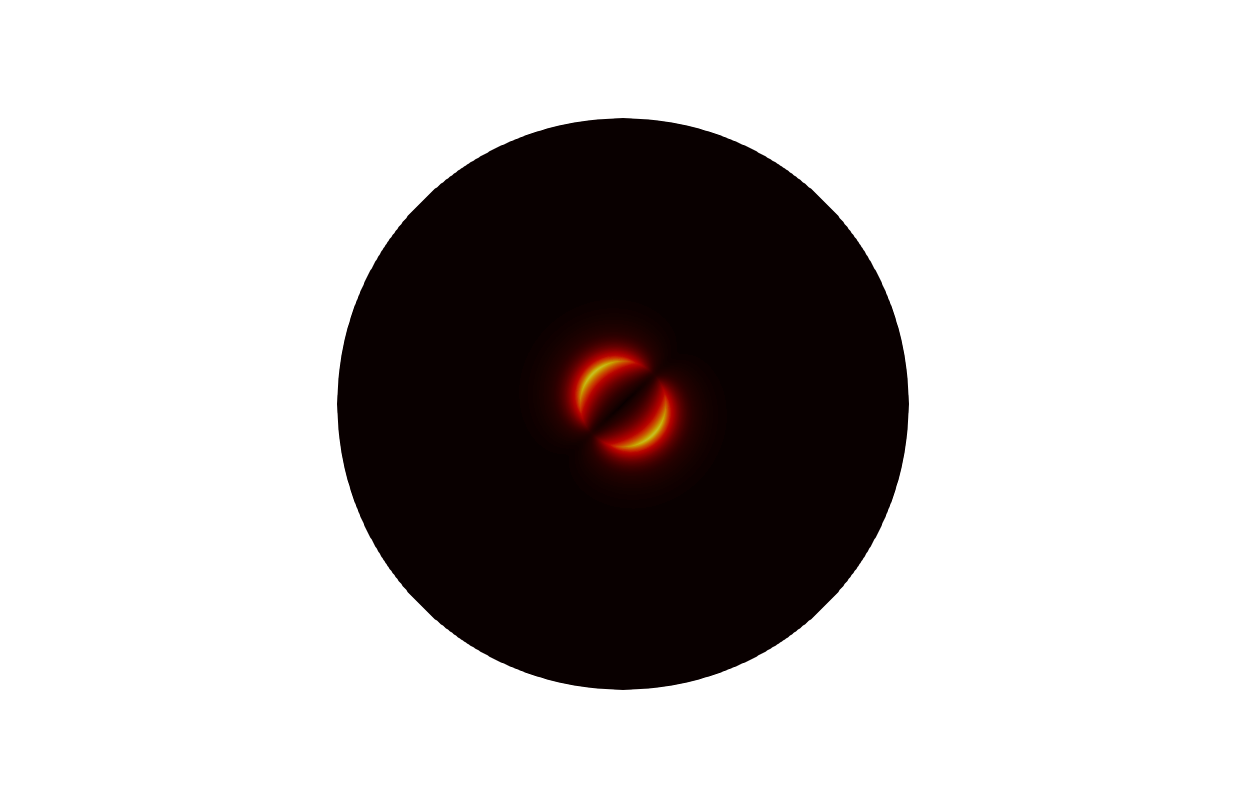}
  }
  \subfloat[$m = 2$]{
    \includegraphics[width=4.2cm,trim=9.9cm 2cm 7.0cm 4cm,clip]{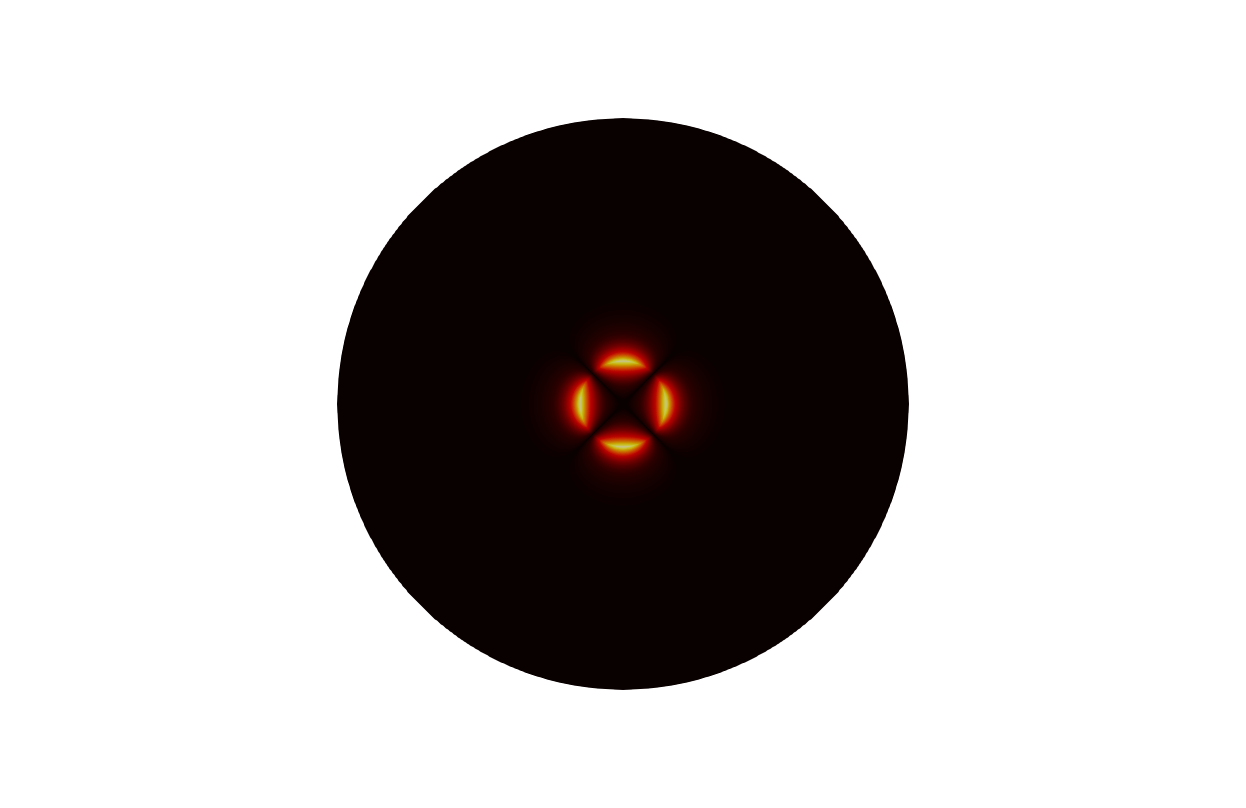}
  }
  \subfloat[$m = 3$]{
    \includegraphics[width=4.2cm,trim=9.9cm 2cm 7.0cm 4cm,clip]{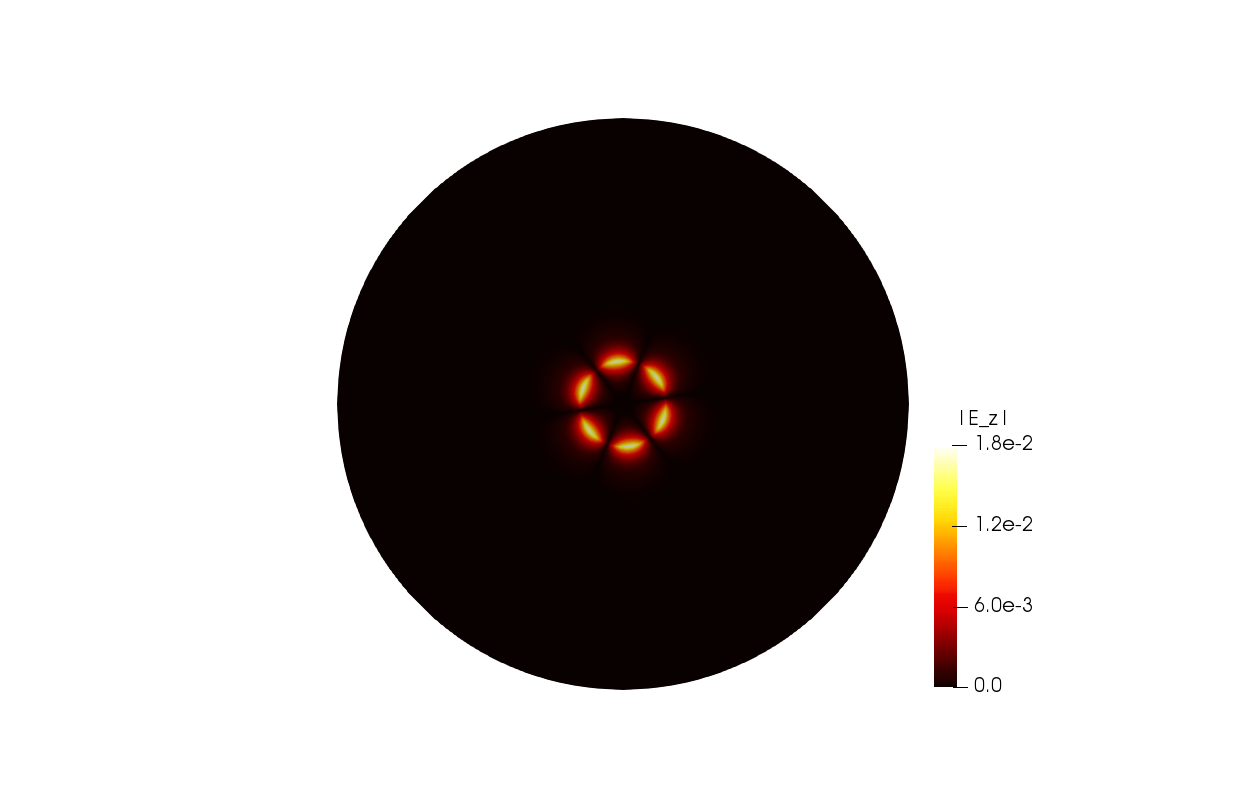}
  }
  \\
  \subfloat[$m = 1$]{\includegraphics{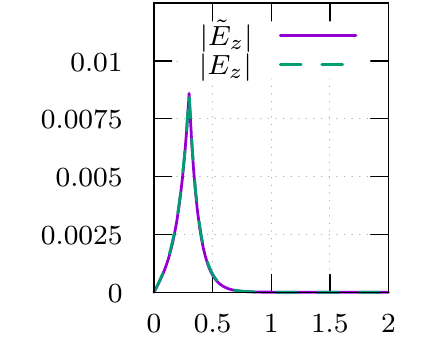}}
  \subfloat[$m = 2$]{\includegraphics{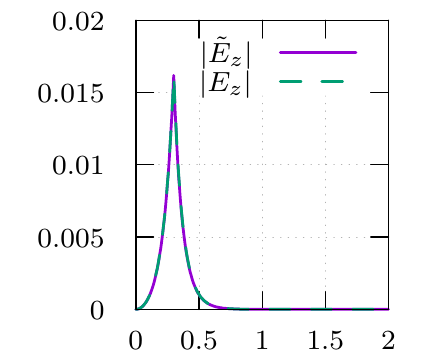}}
  \subfloat[$m = 3$]{\includegraphics{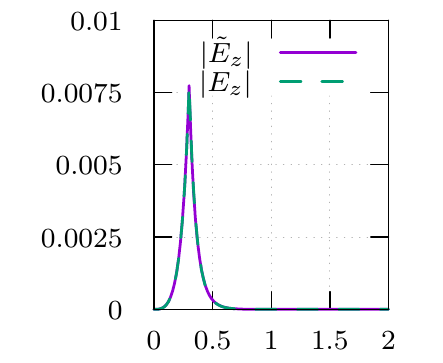}}

  \caption{Computational results for modal order $m = 1,2,3$ for a
    spherical reference configuration with constant material parameters,
    and prescribed conductivity $\ssr = 0.002 + 0.20i$ at radius $\rho_i = 0.3$.
    The computational domain is a disc with radius $R=2.0$.
    (a-c) shows the magnitude of the numerically computed electric
    field, $\tilde E_z$. The computed electric field is compared against an
    analytic solution, $E_z$, in (d)-(f).}
  \label{fig:comparison_figure}
\end{figure}

By assuming constant material parameters, the quartic eigenvalue
problem (\ref{eq:quartic}) does not exhibit any hybridization and reduces to
a linear eigenvalue problem: Find $u\in H^1(\Omega,\mathbb{C})$ s.\,t.
\begin{align}
  \label{eq:linear}
  \tag{L}
  \mathcal{L}(u)[\varphi] := a(u,\varphi) + k_z^2\;m(u,\varphi),
\end{align}
for $\varphi\in H^1(\Omega,\mathbb{C})$ and where we have introduced the bilinear forms
\begin{align}
  \begin{aligned}
  \begin{cases}
    a(u,\varphi) = \int_\Omega \nabla_s u\cdot\nabla_s\overline\varphi\dx
      - \int_\Omega \mur\er u\,\overline\varphi\dx
      - i\int_\Sigma \mur\ssr u\,\overline\varphi\dox, \\[0.5em]
    m(u,\varphi) = \int_\Omega u\,\overline\varphi\dx
      + i\int_\Sigma\ssr\eri u\,\overline\varphi\dox.
    \label{eq:weakform_linear}
  \end{cases}
  \end{aligned}
\end{align}
For a simple spherical geometry that is rotationally invariant, the field
solution can be expressed as a superposition of the modified Bessel
functions of the first and second kind. In the case of a waveguide with a
single circular interface $\Sigma$, i.e., where $\Sigma$ is described by a
circle with origin $0$ and radius $\rho_i$, the analytical solution takes the
following form.
\begin{align}
  E_z & =
  \begin{cases}
    \begin{aligned}
      & A_m I_m(ik_s\rho)e^{im\theta}e^{ik_zz} & \text{for } \rho < \rho_i, \\
      & B_m K_m(ik_s\rho)e^{im\theta}e^{ik_zz} & \text{for } \rho > \rho_i,
    \end{aligned}
  \end{cases}
  \\[0.5em]
  H_z & =
  \begin{cases}
    \begin{aligned}
      & C_m I_m(ik_s\rho)e^{im\theta}e^{ik_zz} & \text{for } \rho < \rho_i, \\
      & D_m K_m(ik_s\rho)e^{im\theta}e^{ik_zz} & \text{for } \rho > \rho_i,
    \end{aligned}
  \end{cases}
\end{align}
where $I_m$ and $K_m$ denote the modified Bessel functions of the first and
second kind, respectively, and $A_m,B_m,C_m$, and $D_m$ are constants that
are determined by the boundary conditions~\cite{gao14a,liu16}. Assuming
that the conducting film is located on the boundary of the interior circle
with radius $\rho_i$, we equate the jump conditions (\ref{eq:jump}) of each
field component. Then (\ref{eq:jump}) reduces to the following algebraic
condition, from which we can retrieve the propagation constant, $k_z$:
\begin{align}
  \det(M-\ssr N) &=0,
  \label{eq:analytical}
\end{align}
where
\begin{align*}
      M &:=
    \begin{pmatrix}
      I_m(ik_s R) & 0 & -K_m(ik_s R) & 0 \\[0.5em]
      \displaystyle\frac{k_zmI_m(ik_s R)}{Rk_s^2} & -\displaystyle\frac{\mu I_m'(ik_sR)}{k_s} &
        -\displaystyle\frac{k_zmK_m(ik_s R)}{Rk_s^2} & \displaystyle\frac{\mu K_m'(ik_sR)}{k_s} \\[0.5em]
     \displaystyle\frac{\epsilon_1 I_m'(ik_sR)}{k_s}  &
        \displaystyle\frac{k_zmI_m(ik_sR)}{Rk_s^2} & -\displaystyle\frac{\epsilon_2 K_m'(ik_sR)}{k_s} &
        -\displaystyle\frac{k_zmK_m(ik_sR)}{Rk_s^2} \\[0.5em]
      0 & I_m(ik_sR)
        & 0 & -K_m(ik_sR)
    \end{pmatrix}
\end{align*}
and
\begin{align*}
      N &:=
    \begin{pmatrix}
      0 & 0 & 0 & 0 \\[0.5em]
      0 & 0 & 0 & 0 \\[0.5em]
      I_m(ik_s R) & 0 & 0 & 0 \\[0.5em]
      -\displaystyle\frac{k_zmI_m(ik_sR)}{Rk_s^2} & \displaystyle\frac{\mu I_m'(ik_sR)}{k_s} & 0 & 0
    \end{pmatrix}.
\end{align*}
We solve for the zeros of \eqref{eq:analytical} numerically via a root
finding algorithm for modal orders $m = 1,2$ and $3$. The computed values
are then compared against those of the linear problem (\ref{eq:linear}) and
of the quartic problem (\ref{eq:quartic}) (see Table~\ref{tab:validation}
and Figure~\ref{fig:comparison_figure}).
\begin{table}[tb]
  \center
  \captionsetup[subfloat]{}
    \footnotesize
    \begin{tabular} {r c c c}
      \toprule
      & \multicolumn{1}{c}{Eigenvalues from (\ref{eq:linear})} &
      \multicolumn{1}{c}{Eigenvalues from (\ref{eq:quartic})} &
      \multicolumn{1}{c}{Eigenvalues from (\ref{eq:analytical})} \\
      \cmidrule(lr){2-2}
      \cmidrule(lr){3-3}
      \cmidrule(lr){4-4}
      Mode  & $k_z$  & $k_z$ & $k_z$
      \\[0.5em]
      1 & $9.447\pm0.090468i$ & $9.447\pm0.090467i$ & $9.447\pm0.090467i$  \\
      2 & $13.00\pm0.090641i$ & $13.00\pm0.090640i$ & $13.00\pm0.090641i$ \\
      3 & $16.17\pm0.099812i$ & $16.17\pm0.099812i$ & $16.17\pm0.099813i$ \\
      \bottomrule
    \end{tabular}
  \caption{
    Validation of computed eigenvalues form the linear problem
    (\ref{eq:weakform_linear}), the quartic problem (\ref{eq:quartic}), and
    the analytical approach (\ref{eq:analytical}) for a single-layer waveguide.
    Note that the values obtained are in agreement with confidence level of less
    than 1\%. Parameters used are
    $R = 2.0, \ssr = 0.002 + 0.20i, \er = \mur\equiv1,$ and $\rho_i = 0.3$.
    }
  \label{tab:validation}
\end{table}
We now validate our numerics with the analytical results. Three validations
are carried out: analytical, numerical with linear eigenvalue, and lastly,
the quartic eigenvalue problem. For simplicity, we assume the computational
domain is isotropic, with material parameters $\er = \mur\equiv1$. A
conducting interface is coated on the boundary of the interior circle with
radius $\rho_i = 0.3$, on a circular domain with radius $R=2.0$. The
surface conductivity is set to $\ssr= 0.002 + 0.20i$.

The analytic computation of the propagation constant $k_z$ requires finding
the complex roots of the determinant of the $4\times4$ matrix. We will
consider modal orders of $m = 1,2$ and $3$ to ease the computation. The
computational results displayed in Table~\ref{tab:validation} deviate by
less than 1\% from each other. We can thus expect a confidence level of 1\%
or better in our numerical computations. Figure~\ref{fig:comparison_figure}
shows the intensity of the numerically computed electric field, $\|E_z\|_2$,
and a comparison of, both, the analytic and numerical solutions. We thus
conclude that our numerical framework is a reliable model that can
effectively simulate hybrid plasmonic modes.


\section{Eigenvalue computations of the linearized companion problem}
\label{sec:results}

In this section, we present a number of computational results obtained from
solving the quartic eigenvalue problem (\ref{eq:quartic}) for a class of
prototypical waveguides with gradient-index materials. We examine
numerically how the spectrum of such a hybridized configuration behaves
under modification of spatially dependent material parameters,
$\er(\rho,\theta)$. We further investigate the relationship between mesh
deformation and the quality factor (defined as the ratio of real part of
the eigenvalue to imaginary part), and study the degree by which the
spectrum changes. All numerical computations are carried out with the
finite element library deal.II~\cite{dealII91}. We use a Krylov-Schur
method to compute solutions of the linearized eigenproblem
\eqref{eq:quadratification}~\citep{slepc05}.

We demonstrate numerically how it is possible to attain an improved quality
factor by prescribing the host material with a radially-varying refractive
index profile. This is methodically carried out in the subsequent
subsections. First, a parameter study is conducted to validate our choice
of discretization parameters. Second, we solve the quartic eigenvalue
problem (\ref{eq:quadratification}) using a number of permittivity
functions, and observe how the spectrums differ from those obtained in an
isotropic medium. Lastly, we deform our computational domain to demonstrate
that our numerical framework is equipped to handle even the most general
configuration. The key idea behind this generalization is to show that we can
manipulate the spectrum by manipulating the shape. We make note of the evolution
of eigenmodes, and how our framework can be used as a basis for shape
optimization of gradient-index waveguides.

The spectrum is computed numerically by using SLEPc~\cite{slepc05}, a
general purpose eigensolver built on top of PETSc~\cite{petsc04}. The
eigensolver provides a number of Krylov-space methods, such as the Arnoldi,
Lanczos, Krylov-Schur, and conjugate-gradient methods. For our purposes, we
make use of the Krylov-Schur method for its faster convergence.


\subsection{Validation of discretization parameters}
\label{subsec:setup}
\begin{table}[t]
  \center
  \captionsetup[subfloat]{}
  \subfloat[
    Vary the number of refinements, $n$.
]{
    \footnotesize
    \begin{tabular} {r r r c r r c}
      \toprule
      & \multicolumn{3}{c}{$s_o = 1.5, R = 1.0, n = 4,\text{DoF}=19138$} &
      \multicolumn{3}{c}{$s_o = 1.5, R = 1.0, n = 6,\text{DoF}=303874$} \\
      \cmidrule(lr){2-4}
      \cmidrule(lr){5-7}
      Mode  & Re$k_z$ & Im$k_z$  & Re$k_z/\text{Im}k_z$ & Re$k_z$  & Im$k_z$    & Re$k_z/\text{Im}k_z$
      \\[0.5em]
      %
      1 & 30.5400 & 0.17646 & 173.070 & 30.4160 & 0.17540 & 173.409  \\
      2 & 40.5520 & 0.21898 & 185.186 & 40.5043 & 0.21881 & 185.112  \\
      %
      \bottomrule
    \end{tabular}
  }

  \subfloat[The control case for the parameter studies.
  ]{
    \footnotesize
    \begin{tabular} {r r r c}
      \toprule
      & \multicolumn{3}{c}{$s_o = 1.5, R = 1.0, n = 5,\text{DoF}=76162$}\\
      \cmidrule(lr){2-4}
      Mode  & Re$k_z$ & Im$k_z$  & Re$k_z/\text{Im}k_z$
      \\[0.5em]
      %
      1 & 30.4160 & 0.17582 & 172.995  \\
      2 & 40.5043 & 0.21881 & 185.112  \\
      %
      \bottomrule
    \end{tabular}
  }

  \subfloat[Vary the domain size, $R$.
]{
    \footnotesize
    \begin{tabular} {r r r c r r c}
      \toprule
      & \multicolumn{3}{c}{$s_o = 1.5, R = 0.5, n = 5,\text{DoF}=76162$} &
      \multicolumn{3}{c}{$s_o = 1.5, R = 2.0, n = 5,\text{DoF}=76162$}\\
      \cmidrule(lr){2-4}
      \cmidrule(lr){5-7}
      Mode  & Re$k_z$ & Im$k_z$  & Re$k_z/\text{Im}k_z$ & Re$k_z$  & Im$k_z$    & Re$k_z/\text{Im}k_z$
      \\[0.5em]
      %
      1 & 30.4242 & 0.17528 & 173.575 & 30.8567 & 0.17331 & 178.043  \\
      2 & 40.4905 & 0.21876 & 185.091 & 40.5043 & 0.21881 & 185.112  \\
      %
      \bottomrule
    \end{tabular}
  }

  \caption{
    Validation of discretization parameters: Parameter study with
    permittivity function $\er(\vx) = 3\cdot\chi_{|\vx| < \rho_i} +
    1\cdot\chi_{\rho_i < |\vx| < R}$. $\mur(\vx)\equiv1$ and $\ssr = 0.002
    + 0.20i$. (b) is the control discretization parameters. (a) differs
    from (b) in the number of refinements, and (c) differs from (b) in the
    size of the domain.}
  \label{tab:parameter_studies}
\end{table}
The computational domain, $\Omega$, is chosen to be the circle with radius
1. A spherical PML is enforced for $\rho > 0.8$. The surface conductivity
$\ssr = 0.002+0.20i$ is chosen that is within a realistic parameter
range~\cite{maier17a} and is located at $\rho_i = 0.3$.
Following~\cite{monk03}, the nonnegative scaling function $s(\rho)$ is
chosen to be
\begin{align}
  s(\rho) = s_0\frac{(\rho - 0.8R)^2}{(R - 0.8R)^2},
\end{align}
where we set the free parameter $s_0$ to be $s_0 = 1.5$ in our
computations. We carry out a parameter study to test the validity and the
sensitivity of discretization parameters. Table~\ref{tab:parameter_studies}
summarizes the parameter study quantitatively. As can be seen, the
eigenmodes comptued are stable with respect to variations of PML strength
$s_0$, the number of initial refinements $n$, and domain sizes $R$. A
spectral transformation is carried out in the form of the M{\"o}bius
transformation, with the parameters chosen to be $a = 1, b = -10, c = 1, d
= 10$. We conclude that $R = 1.0$ and $k = 5$ is a valid choice of
discretization parameters.


\subsection{Gradient-index waveguide}

\begin{figure}[tb]
  \centering
  \includegraphics{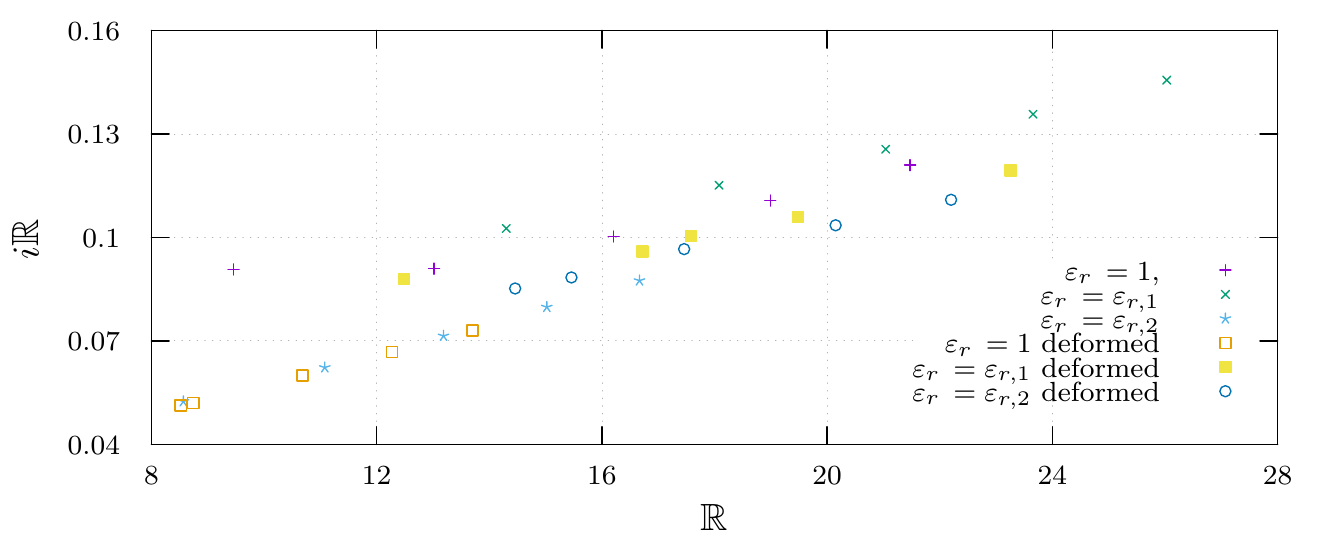}\vspace{-2em}
  \caption{Spectrums computed using different $\er$ for spherical and deformed
  domain. For $\er\equiv1$ and $\varepsilon_{r,1}$, both the real and the
  imaginary parts have decreased after the domain was deformed. The opposite is
  observed for $\varepsilon_{r,2}$. In the spherical domain case, the eigenvalues for
  $\er\equiv1$ and $\varepsilon_{r,1}$ remain relatively close to one another, but
  undergo a noticeable difference after the mesh deformation. The opposite can
  be seen for $\varepsilon_{r,1}$ and $\varepsilon_{r,2}$.
  \label{fig:spectra_original}
    }
\end{figure}

Our numerical framework admits any (locally) differentiable material
parameters. To demonstrate this, we consider the following model material
permittivity functions and analyze their spectrums in relation to those of
the isotropic medium.
\begin{figure}[tb]
  \centering
  \subfloat[
    Surface plot of $\epsilon_{r,1}$
    ]{
    \includegraphics[width=6.0cm,trim=12.0cm 1cm 6.0cm 1cm,clip]{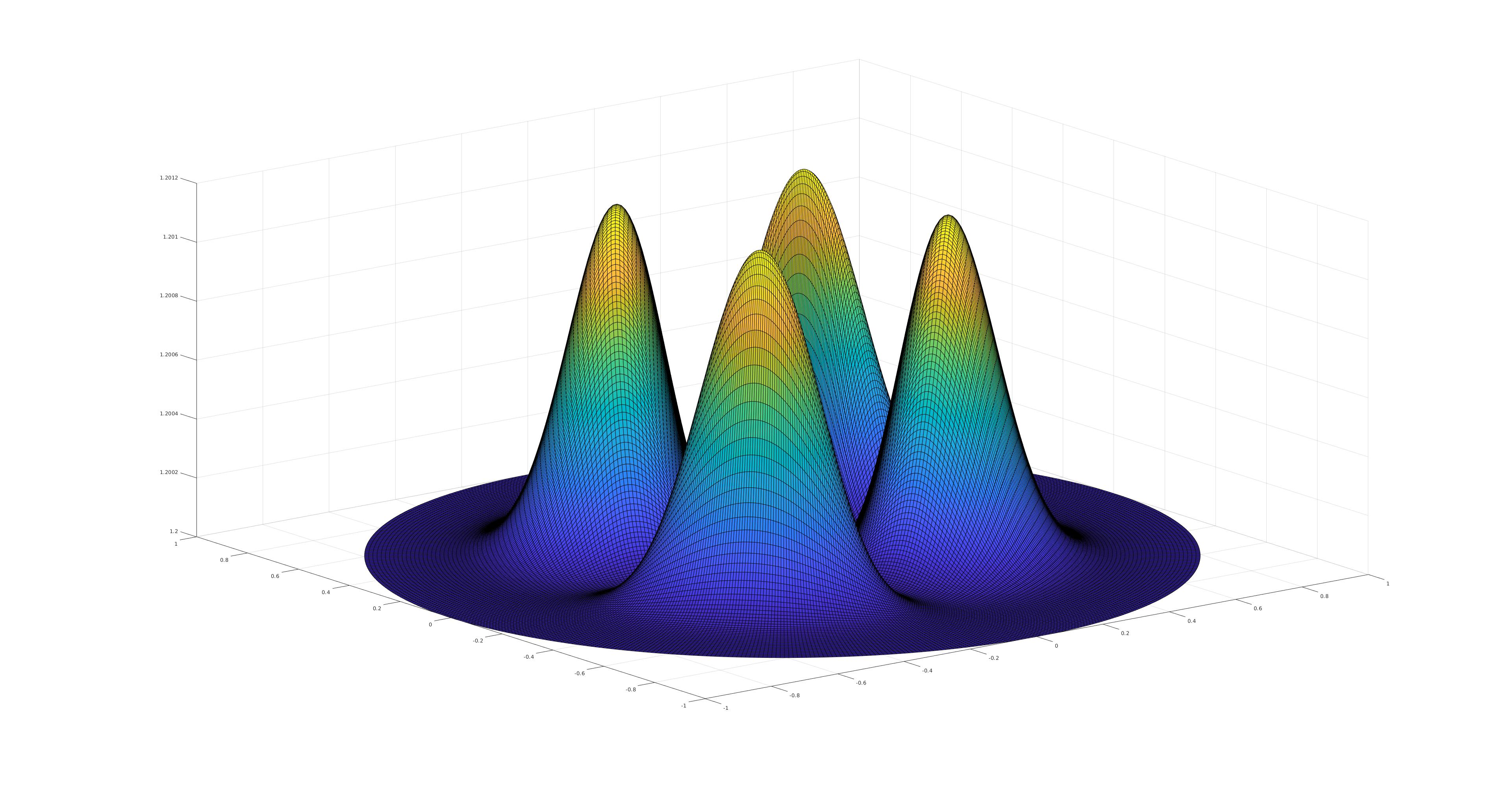}
  }
  \subfloat[
    Surface plot of $\epsilon_{r,2}$
    ]{
    \includegraphics[width=6.0cm,trim=12.0cm 1cm 6.0cm 1cm,clip]{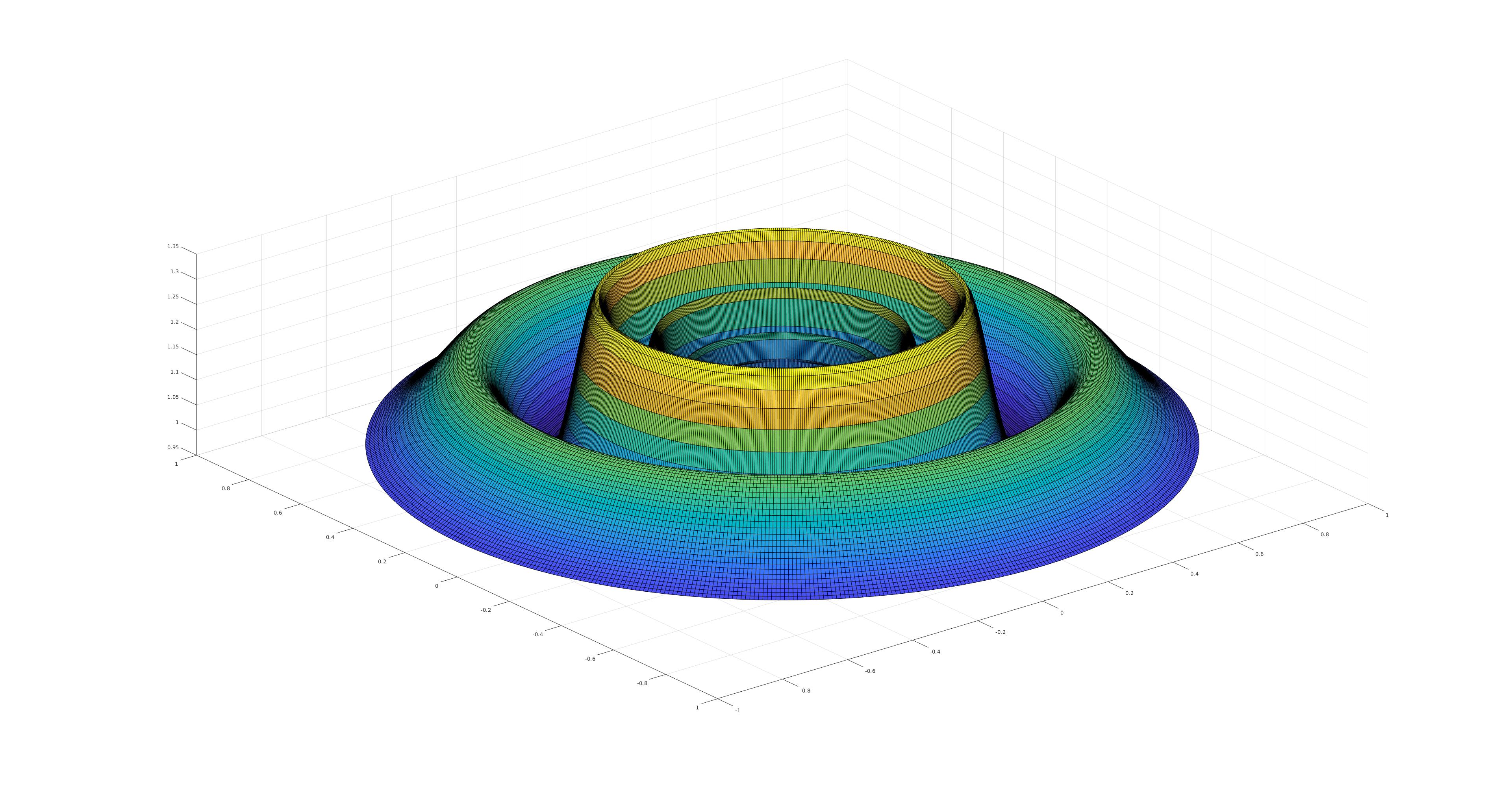}
  }
  \caption{
    A surface plot of the two permittivity profiles used in the
    computations. (a) a 3D plot of $\varepsilon_{r,1}(x,y)$ as given by
    \eqref{eq:permittivity}; (b) a 3D plot of $\varepsilon_{r,2}(x,y)$ as
    given by \eqref{eq:permittivity}.}
  \label{fig:profiles}
\end{figure}
\begin{align}
  \begin{cases}
    \begin{aligned}
      \varepsilon_{r,1}(\rho,\theta) &\;=\; 1.2 + \frac12\sin\left(\frac{\rho^2}{(2\rho_i)^2}
        \sin2\theta\right)\exp\left(-\frac{(\rho-2\rho_i)^2}{(2\rho_i)^2}\right), \\[0.3em]
      \varepsilon_{r,2}(\rho) &\;=\;1+\left(\frac{\rho}{2}+\rho^2\sin\left(\frac{2\pi}{\rho}\right)
        \right)\exp\left(-\frac{(\rho-2\rho_i)^2}{2\rho_i}\right),
    \end{aligned}
  \end{cases}
  \label{eq:permittivity}
\end{align}
where $\rho_i$ is the radius at which the conducting interface is situated
for a circular waveguide. We set $\rho_i = 0.1$ for our computations. The
surface plot of these profiles can be seen in Figure~\ref{fig:profiles}.
The key aspect of these functions is that $\er$ remains constant in the
PML, which enables us to implement the PML as laid out in \ref{subsec:pml}.
The computations are carried out using the unit circular waveguide. For
comparison, we plot the eigenvalues for both isotropic media and materials
defined by (\ref{eq:permittivity}) (see Figure~\ref{fig:spectra_original}).
The quality factor, $\eta = {\text{Re}k_z}/{\text{Im}k_z}$, of the first 5
modes of each of these functions are laid out in
Table~\ref{tab:quality_factor}.

We note of a few observations. The eigenvalues obtained from
$\varepsilon_{r,i}$ are more clustered than those from isotropic media.
Even though the range of $\varepsilon_{r,i}$
($\varepsilon_{r,1}\in(1.0,1.4)$ and $\varepsilon_{r,2}\in(1.0,1.1614)$) is
relatively close to 1, we observe significant changes to the spectrum and
the quality factor. From Table~\ref{tab:quality_factor}, a much longer
propagation is observed for $\varepsilon_{r,2}$ than for
$\varepsilon_{r,1}$, despite their relatively similar range. This
demonstrates that the relationship between $\er$ and the quality
factor, $\eta$, is not trivial, and suggests that it is indeed possible to
improve $\eta$ with a nontrivial gradient-index $\er$.

\begin{table}[tb]
  \center
  \captionsetup[subfloat]{}
  \subfloat[Eigenvalues for spherical domain]{
    \footnotesize
    \begin{tabular} {c r r r r r r r r r}
      \toprule
      & \multicolumn{3}{c}{$\er\equiv1$} &
      \multicolumn{3}{c}{$\er=\varepsilon_{r,1}$} &
      \multicolumn{3}{c}{$\er=\varepsilon_{r,2}$} \\
      \cmidrule(lr){2-4}
      \cmidrule(lr){5-7}
      \cmidrule(lr){8-10}
      Mode  & $\text{Re}k_z$ & $\text{Im}k_z$ & $\eta_3$ & $\text{Re}k_z$ & $\text{Im}k_z$ & $\eta$ & $\text{Re}k_z$ & $\text{Im}k_z$ & $\eta$
      \\[0.5em]
      1 & 13.02 & 0.09093 & 143.2 & 14.30 & 0.1026 & 139.4 & 8.569 & 0.05243 & 163.4 \\
      2 & 16.21 & 0.1003  & 161.7 & 18.08 & 0.1151 & 157.1 & 11.08 & 0.06230 & 177.9 \\
      3 & 18.99 & 0.1107  & 171.6 & 21.04 & 0.1256 & 167.6 & 13.19 & 0.07144 & 184.6 \\
      4 & 21.47 & 0.1210  & 177.6 & 23.66 & 0.1357 & 174.3 & 15.02 & 0.07979 & 188.3 \\
      5 & 23.74 & 0.1309  & 181.4 & 26.03 & 0.1456 & 178.8 & 16.67 & 0.08746 & 190.6 \\
      \bottomrule
    \end{tabular}
  }
  \\[0.1em]
  \subfloat[Eigenvalues for deformed domain]{
    \footnotesize
    \begin{tabular} {c r r r r r r r r r}
      \toprule
      & \multicolumn{3}{c}{$\er\equiv1$} &
      \multicolumn{3}{c}{$\er=\varepsilon_{r,1}$} &
      \multicolumn{3}{c}{$\er=\varepsilon_{r,2}$} \\
      \cmidrule(lr){2-4}
      \cmidrule(lr){5-7}
      \cmidrule(lr){8-10}
      Mode  & $\text{Re}k_z$ & $\text{Im}k_z$ & $\eta_3$ & $\text{Re}k_z$ & $\text{Im}k_z$ & $\eta$ & $\text{Re}k_z$ & $\text{Im}k_z$ & $\eta$
      \\[0.5em]
      1 & 8.521 & 0.05126 & 166.2 & 12.49 & 0.0880 & 141.9 & 14.46 & 0.0852 & 169.8 \\
      2 & 8.745 & 0.05194 & 168.8 & 16.72 & 0.0959 & 163.8 & 15.46 & 0.0884 & 174.9 \\
      3 & 10.69 & 0.06003 & 178.0 & 17.58 & 0.1004 & 175.1 & 17.46 & 0.0966 & 182.1 \\
      4 & 12.27 & 0.06683 & 183.7 & 19.48 & 0.1059 & 184.0 & 20.15 & 0.1035 & 194.8 \\
      5 & 13.71 & 0.07300 & 187.8 & 23.26 & 0.1194 & 194.8 & 22.20 & 0.1109 & 200.2 \\
      \bottomrule
    \end{tabular}
  }
  \caption{Eigenvalues $k_z$ and quality factor
  $\eta=\text{Re}k_z/{\text{Im}k_z}$ for different $\er$ in spherical
  and deformed domains. Noticeable shift in the spectrums is observed from
  (a) to (b), which in turn, has led to non-negligible increases in $\eta$.}
  \label{tab:quality_factor}
\end{table}
%


\subsection{Generalized configuration}
\label{subsec:results}

\begin{figure}[tb]
  \centering
  \subfloat[Deformed mesh]{
    \includegraphics[width=3.0cm,trim=12.0cm 1cm 12.0cm 1cm,clip]{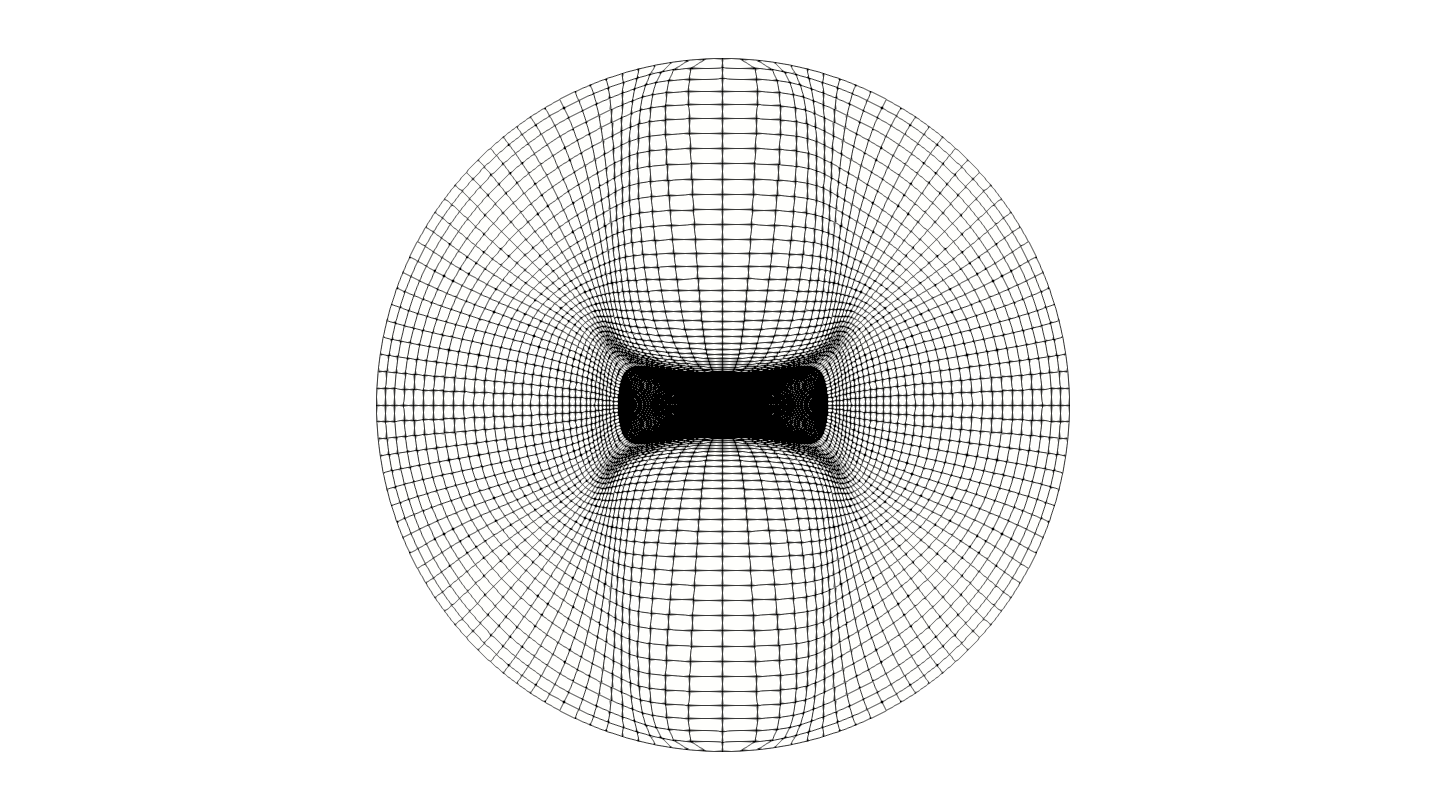}
  }
  \qquad
  \subfloat[zoom]{
    \includegraphics[width=3.0cm,trim=12.0cm 1cm 12.0cm 1cm,clip]{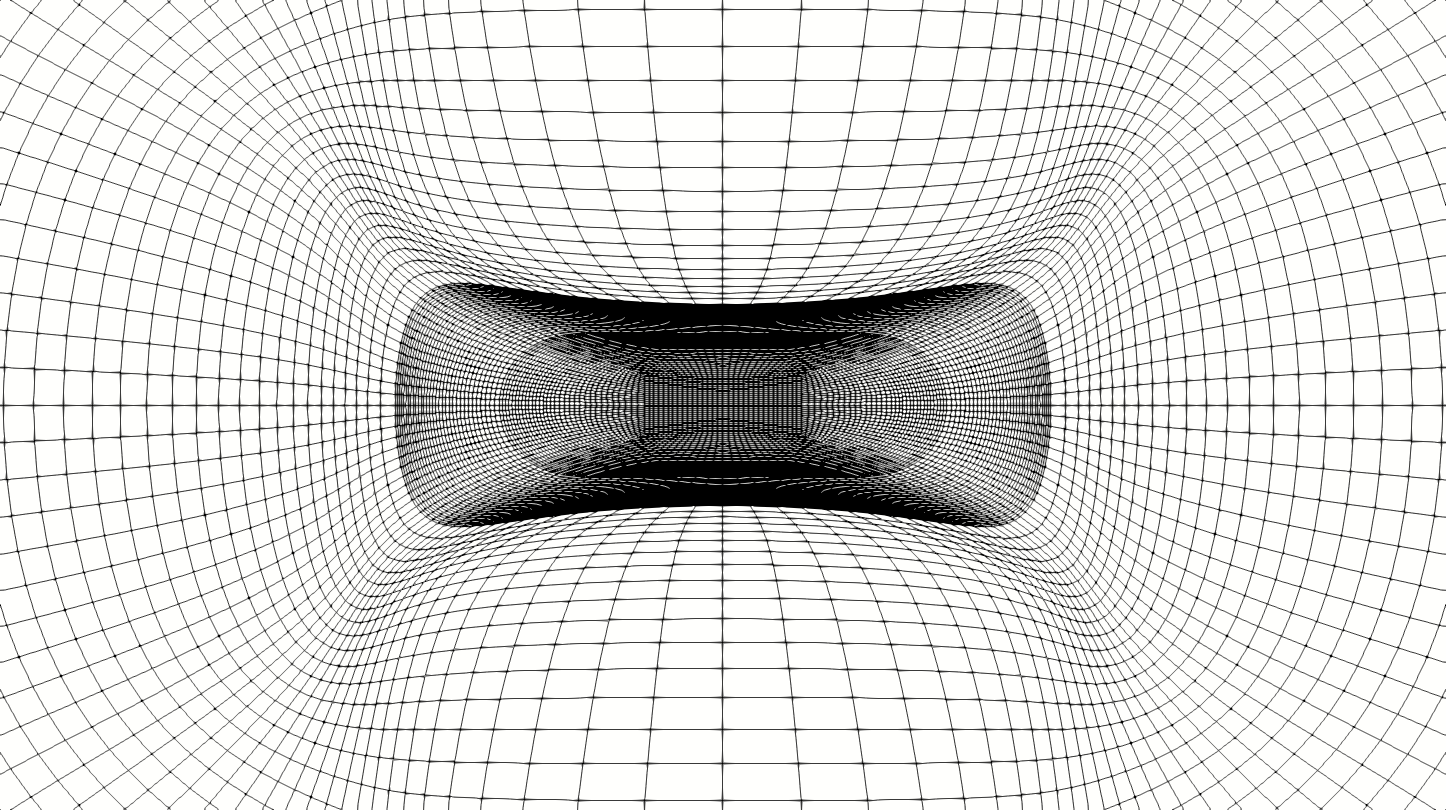}
  }
  \caption{The deformed mesh (a) obtained with the mapping function $\delta$
  outlined below. The mesh has a total number of around
  $370,000$ cells. Additional refinements are a priori enforced around the
  conducting interface.}
  \label{fig:mesh}
\end{figure}
\begin{figure}[tb]
  \centering
  \subfloat[
    ]{
    \includegraphics[width=4.0cm,trim=12.0cm 1cm 6.0cm 1cm,clip]{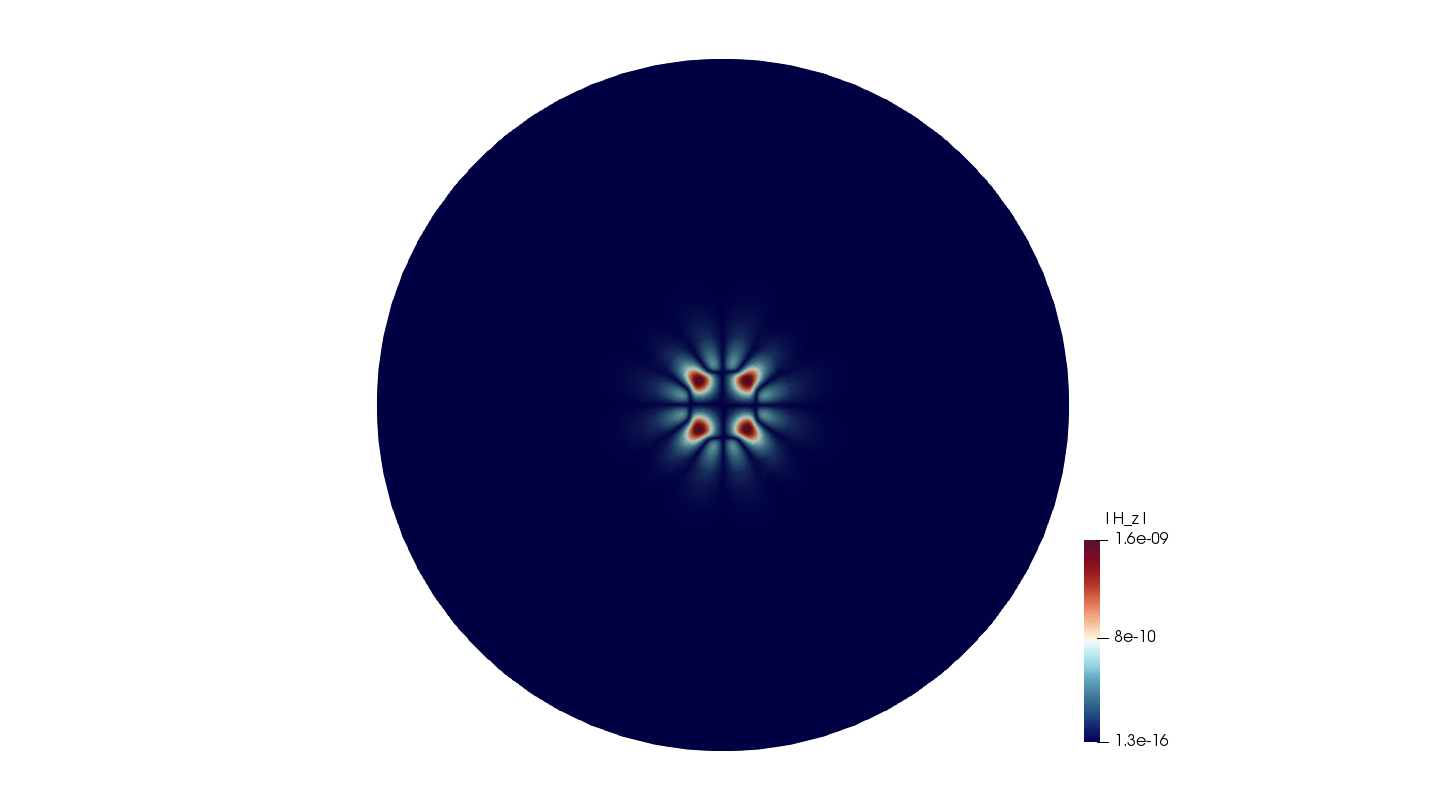}
  }
  \subfloat[
    ]{
    \includegraphics[width=4.0cm,trim=12.0cm 1cm 6.0cm 1cm,clip]{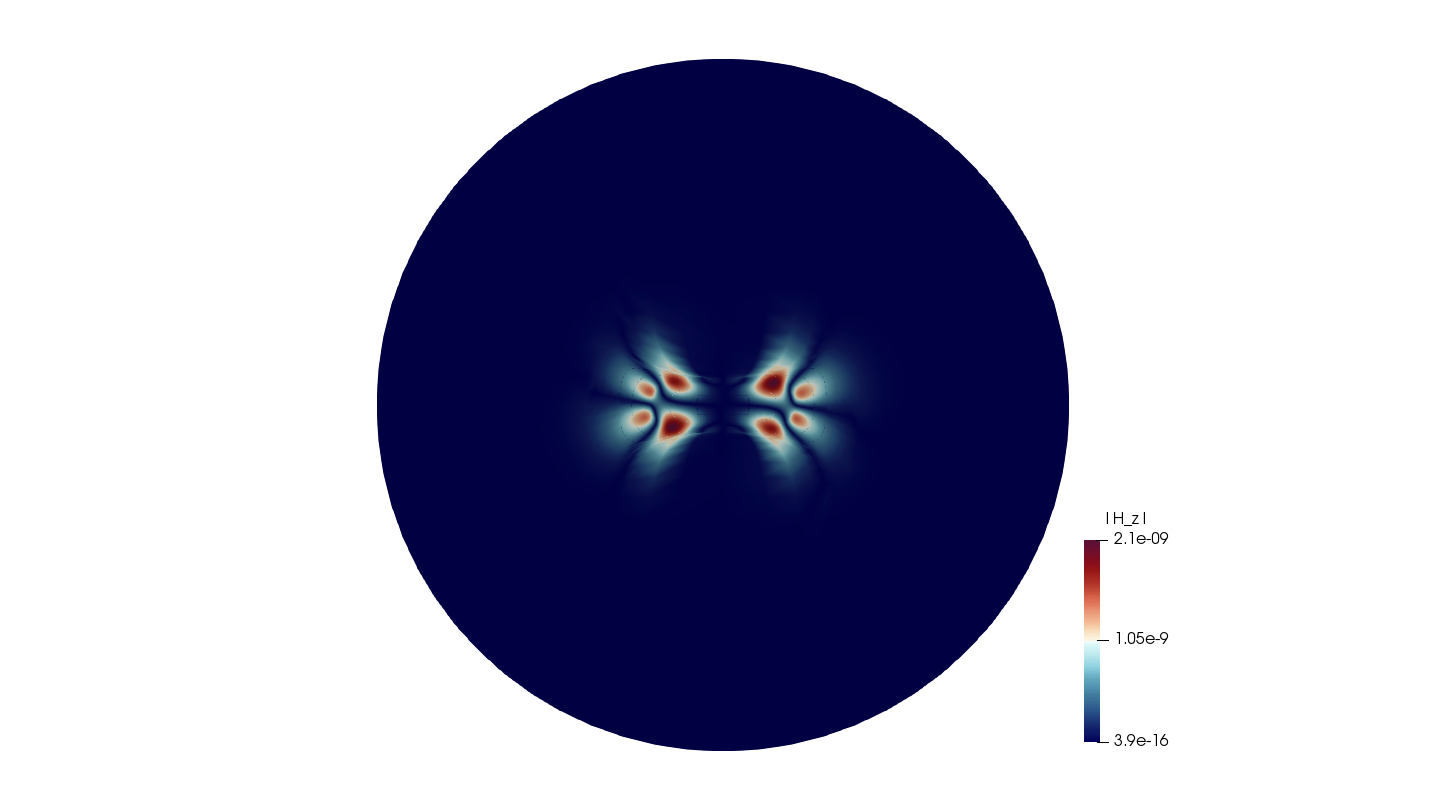}
  }
  \\
  \subfloat[
    ]{
    \includegraphics[width=4.0cm,trim=12.0cm 1cm 6.0cm 1cm,clip]{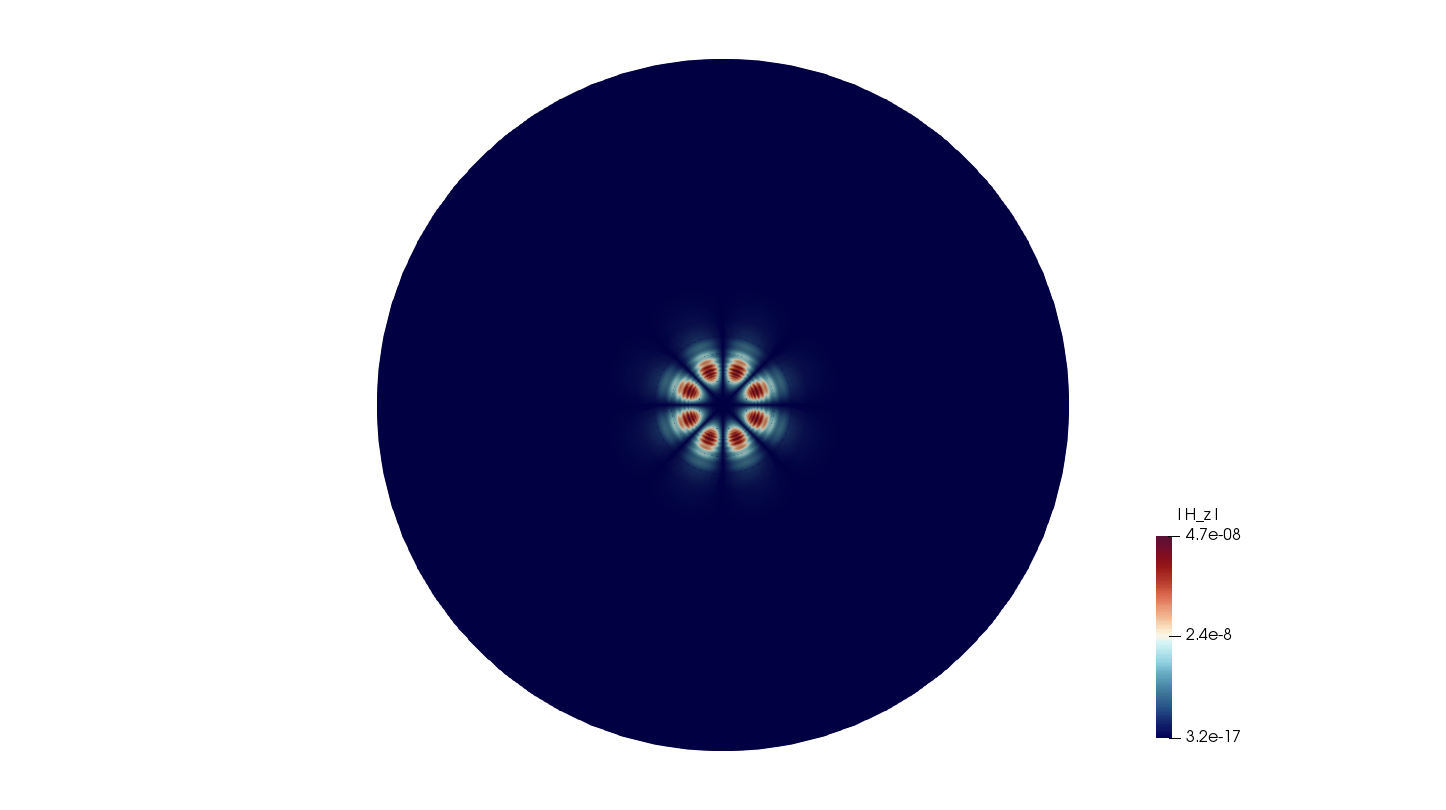}
  }
  \subfloat[
    ]{
    \includegraphics[width=4.0cm,trim=12.0cm 1cm 6.0cm 1cm,clip]{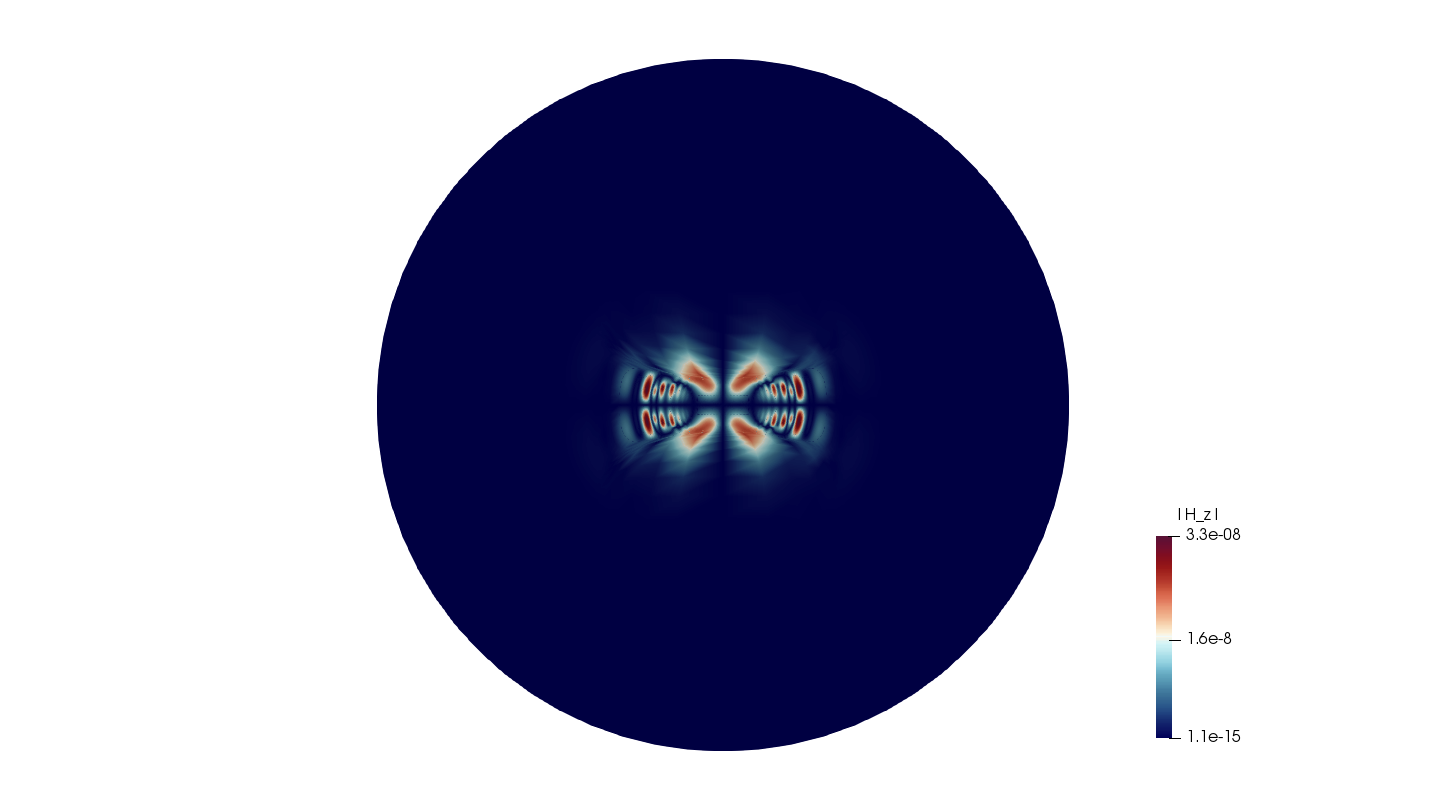}
  }
  \caption{
    An example of $|H_z|$ eigenmode before and after mesh deformation.
    (a) \& (b) are obtained using $\varepsilon_{r,1}$ with modal order of $m=4$
    and (c) \& (d) via $\varepsilon_{r,2}$ with $m=4$.
  }
  \label{fig:mag}
\end{figure}
As a final numerical investigation, we compute a deformed geometric
configuration by using the same permittivity functions introduced in the
preceding section. The purpose of this exercise is two-fold. For one, we
demonstrate that our computational approach can handle large mesh
deformations. Secondly, we demonstrate with this computation that the
quality factor $\eta$ can indeed be controlled and optimized by changing
the shape of the waveguide. This leads to a number of very appealing shape
optimization problems.

Let us introduce a mapping $\delta = (\delta_x, \delta_y)$ that deforms the
mesh near the conducting interface heavily. The only restriction we impose
is that the boundary of the computational domain is circular so as to
preserve the effectiveness of the spherical PML. As a prototypical example
we consider
\begin{align}
  \begin{cases}
    \begin{aligned}
      \delta_x &= 0.8x(1-\arctan^2(4x))\cdot(1+\exp(-|x|))\cdot
        \left(1 - \frac{|x|}{\sqrt{R^2-y^2}}\right)^2, \\[0.3em]
      \delta_y &= \mp A\sin^2(\pi y^2/RT)\cdot(1+\exp(\mp \rho_i y))\cdot
        \left(1 - \frac{|y|}{\sqrt{R^2-x^2}}\right)^2,
    \end{aligned}
  \end{cases}
  \label{eq:mapping}
\end{align}
where $A$ is the displacement amplitude, $T$ is the displacement period,
and $\rho_i$ is the inner radius of the original mesh. For our purpose, we
let $A =55$ and $T=15$. The visualized domain, as defined by
(\ref{eq:mapping}), is shown in Figure~\ref{fig:mesh} and numerical output
can be found in Table~\ref{tab:quality_factor}(b). We anticipate that such
a deformed geometry can be realized in practice by applying stress to a
waveguide on top and bottom deforming the geometry into an almost
rectangular shape. In Figure~\ref{fig:mag}, we plot the magnitude of
hybridized magnetic fields with modal order $m=4$, with $\er =
\varepsilon_{r,1}$ and $\varepsilon_{r,2}$. The conducting interface has
been stretched, which induces a stronger plasmon interaction, and in turn,
an improved quality factor. From Table~\ref{tab:quality_factor}, we observe
that the spectrum can be manipulated by changing the geometry and that the
relationship between the choice of $\er$ and $\eta$ is not trivial. This
presents a potential future research topic for the designing of optical
devices.


\section{Conclusions and Outlook}
\label{sec:conclusion}

In this paper, we formulated a variational framework for the numerical
simulation of guided modes in a waveguide setting with gradient-index host.
This resulted in a quartic eigenvalue problem, which was linearized
via a quadratification approach. The eigenmodes of interest are electromagnetic
SPPs that arise on a conducting closed curve, e.g., graphene-coated waveguide.
The interface is modeled by an idealized, oriented hypersurface.

One of the main advantages offered by our approach is a generalization of
material parameters and geometric configuration. We tested our numerical
treatment of the quartic eigenvalue problem with analytical predictions in
the case of an isotropic medium, and observed excellent agreement. We
assessed the relative strength of computed eigenmodes by quantifying
eigenvalues via the quality factor, and demonstrated using concrete
examples that it is possible to achieve a better quality factor. An
improved quality factor is observed for both gradient-index waveguide and
generalized geometry.

Ideally, we wish to solve the following optimization problem:
\begin{maxi*}
  {\er(\vx),\mur(\vx),\Sigma}{{\text{Re}\,k_z(\er(\vx),\mur(\vx),\Sigma)}\,/\,{\text{Im}\,k_z(\er(\vx),\mur(\vx),\Sigma)}}
  {}{}
  \addConstraint{[\er]_\Sigma = [\mur]_\Sigma = 0, \text{with egularity assumptions}}
  \addConstraint{l(\Sigma) = c, \text{ where } c \text{ is constant.}}
\end{maxi*}
Here, $l(\cdot)$ denotes the length of the curve. In the special case where
$\er\equiv1$ and $\Sigma$ is two infinite parallel layers, the optimization
problem reduces to one discussed in \cite{song19}. We observe that these
generalized constraints can be used as a basis for solving related shape
optimization problems for complicated multilayer optical devices, which is
the subject of future research.


\appendix
\section{Derivation of the weak form}
In this appendix, we carry out in detail the derivation of our weak formulation
(\ref{eq:bilinear_forms}). As a preliminary step, we explain how the longitudinal
component of the guided mode is derived.

\label{app:derivation}
\subsection{Longitudinal component}

The transverse and the longitudinal components of the rescaled time-harmonic
Maxwell's equations with $e^{ik_zz}$ dependence are
\begin{align}
  &\begin{cases}
   \begin{aligned}
     i\mur\vH_s & = \nabla_s\times\hat{z}E_z + ik_z\hat{z}\times\vE_s, \\
     -i\er\vE_s & = \nabla_s\times\hat{z}H_z + ik_z\hat{z}\times\vH_s,
   \end{aligned}
   \label{eq:rescaled_scomponent}
  \end{cases} \\[0.5em]
  &\begin{cases}
   \begin{aligned}
     \nabla_s\times\vE_s & = i\mur\hat{z}H_z, \\
     \nabla_s\times\vH_s & = -i\er\hat{z}E_z.
   \end{aligned}
   \label{eq:rescaled_zcomponent}
  \end{cases}
\end{align}
and the corresponding jump conditions at an interface $\Sigma$ are
\begin{align}
 \begin{cases}
   \begin{aligned}
     \left[\vn\times(\vH_s + \hat{z}H_z)\right]_\Sigma & = \ssr\Big\{
       \big(\vn\times(\vE_s + \hat{z}E_z)\big)\times\vn\Big\}_\Sigma, \\
     \left[\vn\cdot\mur(\vH_s + \hat{z}H_z)\right]_\Sigma & =
     [\vn\times(\vE_s + \hat{z}E_z)]_\Sigma =
     [\vn\cdot\er(\vE_s + \hat{z}E_z)]_\Sigma = 0,
   \end{aligned}
 \end{cases}
 \label{eq:rescaled_jump}
\end{align}
where $\vn$ is the normal vector at $\Sigma$. Equate each component of
(\ref{eq:rescaled_jump}) to obtain
\begin{align}
 \begin{cases}
   \begin{aligned}
     [\left(\vH_s\cdot\vt\right)\hat{z}]_\Sigma & =
       \ssr\Big\{\left(-E_z\vt\right)\times\vn\Big\}_\Sigma
       = \ssr E_z\hat{z}\Big|_\Sigma, \\[0.5em]
     [-H_z\vt]_\Sigma & =
       \ssr\Big\{\left(\vE_s\cdot\vt\right)\hat{z}\times\vn\Big\}_\Sigma
       = \ssr\Big(\vE_s\cdot\vt\Big)\vt\Big|_\Sigma, \\[0.5em]
     [\er \vE_s]_\Sigma\cdot\vn & = [\mur\vH_s]_\Sigma\cdot\vn=
       [\vE_s]_\Sigma\cdot\vt = [E_z]_\Sigma = 0.
   \end{aligned}
 \end{cases}
\end{align}

Substitute one of (\ref{eq:rescaled_scomponent}) into the other to obtain
(\ref{eq:rescaled_transverse}).
\begin{align}
  \begin{cases}
    \begin{aligned}
      k_s^2\vE_s & = i\left(k_z\nabla_s E_z+\mur\nabla_s
      \times H_z\right), \\[0.2em]
      k_s^2\vH_s & = i\left(k_z\nabla_s H_z-\er\nabla_s
      \times E_z\right).
    \end{aligned}
  \end{cases}
  \label{eq:rescaled_transverse}
\end{align}
The second-order time-harmonic Maxwell's equations are
\begin{align}
 \begin{aligned}
   (\nabla_s +
     ik_z\hat{z})\times(\muri(\nabla_s+ik_z\hat{z})\times\vE)-
     \er\vE &\;=\; 0, \\
   (\nabla_s +
     ik_z\hat{z})\times(\eri(\nabla_s+ik_z\hat{z})\times\vH)-
     \mur\vH &\;=\; 0, \\
 \end{aligned}
 \label{eq:secondorder_unscaled}
\end{align}
Equate the $z$-component of (\ref{eq:secondorder_unscaled}) to obtain
(\ref{eq:secondorder_rescaled}).

\subsection{Derivation of the weak form}
We multiply (\ref{eq:rescaled_z}) by $k_s^4$ and distribute it in a particular
manner that eases the handling of the inhomogeneities.
\begin{align}
  \begin{cases}
    \begin{aligned}
      -k_s^4\Big(\Big(\nabla_s\frac1{k_s^2}\Big)\cdot
        \Big(&\er\nabla_s E_z\Big)+\frac1{k_s^2}\nabla_s\cdot
        \Big(\er\nabla_s E_z\Big)\Big) \\
      &- k_zk_s^4\Big(\Big(\nabla_s\frac1{k_s^2}\Big)\cdot
        \Big(\nabla_s\times\hat{z}H_z\Big)\Big) - \er k_s^4 E_z
        \;=\;0
        \\
      -k_s^4\Big(\Big(\nabla_s\frac1{k_s^2}\Big)\cdot
        \big(&\mur\nabla_s H_z\Big)+\frac1{k_s^2}\nabla_s\cdot
        \big(\mur\nabla_s H_z\Big)\Big) \\
      &+ k_zk_s^4\Big(\Big(\nabla_s\frac1{k_s^2}\Big)\cdot
        \Big(\nabla_s\times\hat{z}E_z\Big)\Big) - \mur k_s^4 H_z
        \;=\; 0.
    \end{aligned}
  \end{cases}
  \label{eq:inhomogeneous_ugly}
\end{align}
Since $\nabla_s k_s^{-2} = -k_s^{-4} \nabla_s k_s^2$, some algebra shows
that the expression in the first line of (\ref{eq:inhomogeneous_ugly}) is
equivalent to
\begin{align}
  \begin{aligned}
    -k_s^4 & \left(\left(\nabla_s\frac1{k_s^2}\right)\cdot
    \left(\er\nabla_s E_z\right)+\frac1{k_s^2}\nabla_s\cdot
    \left(\er\nabla_s E_z\right)\right) \\
    & = \nabla_s k_s^2\cdot\left(\er\nabla_s E_z\right)
    - k_s^2\nabla_s\cdot\left(\er\nabla_s E_z\right) \\
    & = \nabla_s\cdot\left(k_s^2
    \left(\er\nabla_s E_z\right)\right) - 2k_s^2
    \nabla_s\cdot\left(\er\nabla E_z\right).
  \end{aligned}
\end{align}
Additionally,
we note that the curl terms in (\ref{eq:inhomogeneous_ugly}) can be written as
\begin{multline}
    - k_z k_s^4\left(\left(\nabla_s\frac1{k_s^2}\right)\cdot
      \left(\nabla_s\times\hat{z}H_z\right)+\frac1{k_s^2}\nabla_s\cdot(\nabla_s\times\hat{z}H_z)\right) \\
    = k_z\nabla_s k_s^2\cdot(\nabla_s\times\hat{z}H_z) -k_zk_s^2\nabla_s\cdot(\nabla_s\times\hat{z}H_z)\\
    = k_z\nabla_s\cdot\left(k_s^2
      \left(\nabla_s\times\hat{z}H_z\right)\right) - 2k_zk_s^2\nabla_s\cdot(\nabla_s\times\hat{z}H_z).
\end{multline}
Even though the last term vanishes, we keep it, as it will be later utilized to
express the interface contribution nicely.

We now test with smooth functions $\varphi$ and $\psi$, and integrate by parts
to arrive at
\begin{multline}
  -(k_s^2\er\nabla_s E_z,\nabla_s\varphi)
      +2(\er\nabla_s E_z,\nabla_s(\overline{k}_s^2\varphi))
      -k_z(k_s^2\nabla_s\times\hat{z}H_z,\nabla_s\varphi) \\
      +2k_z(\nabla_s\times\hat{z}H_z,\nabla_s(\overline{k}_s^2\varphi))
      -(\er k_s^4 E_z,\varphi) \\
  -(k_s^2\mur\nabla_s H_z,\nabla_s\psi)
      +2(\mur\nabla_s H_z,\nabla_s(\overline{k}_s^2\psi))
      +k_z(k_s^2\nabla_s\times\hat{z}E_z,\nabla_s\psi) \\
      -2k_z(\nabla_s\times\hat{z}E_z,\nabla_s(\overline{k}_s^2\psi))
      -(\mur k_s^4 H_z,\psi) \\
  +\langle[k_s^2\er\pn E_z+k_s^2k_z\pt H_z]_\Sigma,\varphi\rangle_\Sigma
  +\langle[k_s^2\mur\pn H_z-k_s^2k_z\pt E_z]_\Sigma,\psi\rangle_\Sigma.
  \label{eq:weakform_app}
\end{multline}
Using (\ref{eq:jump}), \eqref{eq:assumptions}, and (\ref{eq:rescaled_transverse}),
the interface contributions simplify to
\begin{multline}
    \langle[k_s^2(\er\pn E_z+k_z\pt H_z)]_\Sigma, \varphi\rangle_\Sigma +
    \langle[k_s^2(\mur\pn H_z-k_z\pt E_z)]_\Sigma,\psi\rangle_\Sigma \\
      = -i\langle[k_s^4 \vH_\tau]_\Sigma,\varphi\rangle_\Sigma -
      i \langle[k_s^4 \vE_\tau]_\Sigma,\psi\rangle_\Sigma
    = -i\ssr\langle k_s^4 E_z,\varphi\rangle_\Sigma.
    \label{eq:interface}
\end{multline}
Expand the $\nabla_s(\overline{k}_s^2\varphi)$ and
$\nabla_s(\overline{k}_s^2\psi)$ to arrive at
\begin{multline}
  (\mur\er^2\nabla_s E_z,\nabla_s\varphi)+2(\er\nabla_s E_z,\nabla_s(\oer\omur)\varphi) \\
      +k_z(\mur\er\nabla_s\times\hat{z}H_z,\nabla_s\varphi)
      -k_z^2(\er\nabla_s E_z,\nabla_s\varphi) \\
      +2k_z(\nabla_s\times\hat{z}H_z,\nabla_s(\oer\omur)\varphi)
      -k_z^3(\nabla_s\times\hat{z}H_z,\nabla_s\varphi) - (\er k_s^4 E_z,\varphi) \\
  + (\er\mur^2\nabla_s H_z,\nabla_s\psi)+2(\mur\nabla_s H_z,\nabla_s(\oer\omur)\psi) \\
      -k_z(\mur\er\nabla_s\times\hat{z}E_z,\nabla_s\psi)
      -k_z^2(\mur\nabla_s H_z,\nabla_s\psi) \\
      -2k_z(\nabla_s\times\hat{z}E_z,\nabla_s(\oer\omur)\psi)
      +k_z^3(\nabla_s\times\hat{z}E_z,\nabla_s\psi) -(\mur k_s^4 H_z,\psi) \\
  -i\ssr\langle k_s^4 E_z,\varphi\rangle_\Sigma = 0.
\end{multline}
%


\end{document}